\documentclass[11pt,a4paper]{article}
\usepackage{amsfonts}
\usepackage{}
\usepackage{jheppub}

\title{Spectral function and quark diffusion constant in non-critical holographic QCD}

\author{Yan Yan Bu,}
\author{Jin Min Yang}

\affiliation{Institute of Theoretical Physics, Academia Sinica, Beijing 100190, China}

\emailAdd{yybu@itp.ac.cn}
\emailAdd{jmyang@itp.ac.cn}

\abstract{Motivated by recent studies of intersecting D-brane systems in critical string theory and phenomenological AdS/QCD models, we present a detailed analysis for the vector and scalar fluctuations in a non-critical holographic QCD model in the high temperature phase, i.e., the chiral symmetric phase. This model is described by $N_f$ pairs of D4 and $\overline{\text{D4}}$ probe branes in a non-critical ${AdS_6}$ black hole background. Focusing on the hydrodynamic as well as the high frequency limit, we analytically obtain spectral functions for vector and scalar modes on the flavor probe. Then we extract the light quark diffusion constant for flavor current using three different methods and find that different methods give the same results. We also compute the heavy quark diffusion constant for comparison with the light quark case.}


\begin{document}
\maketitle
\flushbottom
\section{Introduction}\label{introduction}
Gauge/gravity duality \cite{Maldacena,Gubser,Witten,Maldacena1} has become a powerful tool in studying strongly coupled non-Abelian gauge theories. Since its invention, it has been expected to be helpful in understanding some mysterious behaviors of low energy QCD. Although the gravity dual of the realistic QCD has not yet been established till now, the Sakai-Sugimoto model \cite{Sakai,Sakai1} has successfully realized certain aspects of low energy phenomena of QCD: the confinement/deconfinement transition has been recognized as the Hawking-Page phase transition \cite{Witten1}, and the non-Abelian chiral symmetry breaking/restoring phase transition also has a geometrical realization \cite{Sakai,Aharony,Parnachev}. However, this model has the undesired KK modes because it is based on critical string theory and has the unwanted internal space. What is worse is that these KK modes and hadrons have the same mass scale and there is no way to disentangle the two scales. One effective way to overcome this serious problem is to consider string theory in non-critical dimensions. Some non-critical string models of QCD have been constructed in \cite{Polyakov,Kuperstein,Kuperstein1,Klebanov,Casero,Armoni,
Giveon,Ferretti,Ferretti1, Alvarez,Ghoroku,Bertoldi,Bigazzi,Ashok1,Fotopoulos,Israel,
Lugo,Murthy,Niarchos,Bigazzi1,Ashok2,Gursoy}.

In this paper we consider a model of non-critical ${AdS_6}$ black hole solution \cite{Kuperstein,Kuperstein1,Casero} (corresponding to the deconfined phase), in which the flavor quarks can be introduced by adding D4 and $\overline{\text{D4}}$ probe branes \cite{Karch,Karch1} into this black hole background. As the non-critical version of the Sakai-Sugimoto model, this model has also reproduced many properties of low energy QCD. Some studies about this model have been performed in the literature, e.g., the thermal aspects\footnote{Thermodynamics and probe energy loss in similar non-critical ${AdS_5}$ plasma have been studied in \cite{Bertoldi1}.} have been studied in \cite{Mazu} and the chiral phase transition with external electromagnetic fields and chemical potential have been examined in \cite{Cui,Cui1,Davody}. The chiral phase transition can be represented by the embedding profile of flavor branes: the U-shape configuration stands for the chiral breaking phase while the parallel branes and anti-branes states the chiral symmetric phase \cite{Mazu}. The hadron physics related to low energy QCD have been intensively studied in \cite{Kuperstein1,Casero,Mazu,Mintakevich}.

One motivation for studying such a non-critical holographic QCD model lies in that this model is different from the well-studied intersecting D-brane systems in critical string theory and some phenomenological AdS/QCD models which lack theoretical completions. Special features of this model are that the dilaton is a constant, which is same as the well-studied D3 brane geometry, and supersymmetry is completely broken, which is same as the Sakai-Sugimoto model. Given that studies on hadron physics and chiral phase transition in this model have reproduced many similar features to its critical version, we here focus on the non-equilibrium state physics (not far from equilibrium state) of quark sector---the retarded Green's function of mesonic operator and the corresponding spectral functions. We here mainly probe some differences and similarities between the Sakai-Sugimoto model and its non-critical version. Meanwhile, one byproduct of this study is to further demonstrate some universal features of gauge/gravity duality as well as eliminate some non-universal ones which are peculiar to some models.

Specifically, we in this work study the vector/scalar fluctuations on flavor probe, calculate their correlation functions\footnote{Studies about retarded Green's functions under the gauge/gravity framework have gained much attention, for reviews see \cite{Son,Skenderis}.} to reproduce the meson spectral functions and then extract quark diffusion constant. We here work in the quenched approximation, i.e., we do not take into account the backreation of flavor probe branes on the $AdS_6$ black hole background. For simplification we will focus on the chiral symmetric phase which is also referred to as high temperature phase. In this phase mesons should not be confined or bounded together, moreover, they are no longer long-lived in contrast to mesons in the chiral broken phase. Oppositely, we expect them to be quasi-normal modes in the black hole backgrounds. This phenomenon can be described as meson melting \cite{Peeters,Hoyos} in chiral symmetric phase (actually, this idea is subtle when considering the processes of chiral phase transition because mesons prior to transition do not correctly correspond to quasi-normal modes after transition \cite{Paredes}). In general, such a study usually needs too much numerical work and, as a result, applicability of results depends on the precision of numerical computation. So in this paper we will merely consider the hydrodynamic as well as high frequency limit, both of which allow for analytical calculations.

In the hydrodynamic limit, the frequency and spatial momentum are much smaller than the temperature of the holographic system, and therefore the system reduces to hydrodynamic and we can consequently probe its hydrodynamic properties. Using gauge/gravity duality to study hydrodynamic properties of strongly coupled non-Abelian gauge theories has gained great success \cite{Son1,Policastro,Kovtun,Kovtun1,Kovtun2,Teaney,Nunez,Son2,Starinets,Iqbal,
Benincasa,Myers,Myers2,Policastro2,Herzog1,Herzog3,Buchel,Benincasa1,Buchel1,Benincasa2,
Mas3,Bigazzi2,Bigazzi3,Gursoy2} (for reviews see \cite{Son,Casalderrey-Solana}). In our study we will use the method proposed in \cite{Son1} to get the hydrodynamic limit for meson spectral functions, whose form is in agreement with expectation from boundary thermal field theory, and then we extract the quark diffusion constant from the correlation function in the diffusive channel. To check correctness of such an extraction, we use another two different ways to compute this diffusion constant and find that three different methods give same results. This confirms the universality of the membrane paradigm, which is proposed in \cite{Kovtun} and further developed in \cite{Starinets,Iqbal} for study of transport coefficients in strongly coupled thermal field theories from gauge/gravity duality approach. Because the quark introduced by probe D-branes is light (here, precisely speaking, it is massless), we refer to this quark diffusion as the light quark diffusion which is different from the heavy quark case being introduced later.

We will also study the high frequency limit using the WKB approximation. In this regime, we analytically obtain spectral functions for different modes. Their behaviors are not the same: for vector modes they are scaling as $\omega$$^{2}$ which is the same as previous results \cite{Myers}, while for the scalar one it scales as $\omega$$^{6}$ which has not been observed in previous studies. Actually, these scaling behaviors can also be extracted from dimensional analysis for dual operators.

For completeness, we then calculate the heavy quark diffusion constant in this model. The heavy quark can be modeled by an open string stretching from the boundary to the bulk. This configuration has been used in studying jet quenched parameter \cite{Liu,Liu1,Argyres}, drag force \cite{Gubser2} and energy loss \cite{Herzog,Herzog2} for heavy quark moving through quark-gluon plasma (for reviews, see \cite{Gubser1,Casalderrey-Solana}). We here just take into account a static string and extract the diffusion constant from the correlation function of the string's embedding
coordinate, which is proposed in \cite{Casalderrey-Solana1} and studied in \cite{Casalderrey-Solana1,Gubser3,Casalderrey-Solana2} for $\mathcal{N}=4$ super-Yang-Mills plasma. The same calculation for the Sakai-Sugimoto model was done in \cite{Pang}. Our result is similar to that of \cite{Casalderrey-Solana1} but different
from \cite{Pang}. A recent elaborate study for heavy quark diffusion in a non-conformal background can be found in \cite{Gursoy1}.

The remainder of this paper is organized as follows. In section \ref{section2}, we set up backgrounds for our study, introducing the holographic model and notation conventions, calculating the charge susceptibility for later convenience and working out equations of motion for flavor vector and scalar fluctuations. Section~{\ref{sec:low}} is the main part of our study, in which we intensively analyze the fluctuations under the hydrodynamic limit and extract the light quark diffusion constant in three ways. Section~{\ref{sec:high}} is concentrated on the study of spectral functions in the high frequency limit using WKB method. In section~{\ref{section6}} we turn to the calculation of the heavy quark diffusion constant. The final section gives some discussions and conclusions as well as some open questions.

\section{Holographic setup and notation conventions}
\label{section2}
\subsection{Non-critical ${AdS_6}$ black hole background}\label{setup}
We take ${N_c}$ coincident D4-branes wrapped on a circle in six dimensions. If we take the large ${N_c}$ and near horizon limits for this brane system, they should be described by classical solutions to the type IIA supergravity in six dimensions. To reproduce the confinement/deconfinement phase transition, the color branes have been chosen to wrap one circle and we will impose periodic boundary conditions for bosons and antiperiodic ones
for fermions along the wrapped circle to explicitly break supersymmetry \cite{Witten1}. Then the gravitational background is dual to five dimensional gauge theory and when energy scale is much smaller than inversion of the circle radius it is effectively four dimensional. Due to the antiperiodic boundary conditions imposed for fermions, the left freedom in the low energy case is just the adjoint gluons. Thus we are left with a non-supersymmetric pure gauge theory with gauge group SU(${N_c}$) in the low energy limit. To describe quarks in the fundamental representation of the gauge group, one introduces ${N_f}$ pairs of D4 and $\overline{\text{D4}}$ branes into this background following \cite{Karch,Karch1}. As mentioned in the introduction, in this work we will take the assumption that $N_f\ll N_c$ and then the backreation of flavor probe branes on the $AdS_6$ black hole background can be ignored, i.e., we do the calculations under the quenched approximation. As in the Sakai-Sugimoto model, the flavor probe is transverse to the wrapped circle and branes configuration has been summarized in Table~\ref{table}.
\begin{table}[!h]
\tabcolsep 0pt
\caption{Branes configuration}
\vspace*{-12pt}
\begin{center}
\def\temptablewidth{0.5\textwidth}
{\rule{\temptablewidth}{1pt}}
\begin{tabular*}{\temptablewidth}{@{\extracolsep{\fill}}ccccccc}
& $t$ &$x^1$ &$x^2$ &$x^3$ &$x^4$ &$u$ \\   \hline
       $\text{color}$ D4& $\times$ &$\times$&$\times$&$\times$&$\times$&   \\
       $\text{flavor}$ $\text{D4}/\overline{\text{D4}}$&$\times$&$\times$&$\times$&$\times$&&$\times$  \label{table}\\
       \end{tabular*}
       {\rule{\temptablewidth}{1pt}}
       \end{center}
       \end{table}

Now we directly write down the background geometry for the deconfined phase using the notation conventions of \cite{Mazu}:
\begin{eqnarray}
ds^2&=&\left( \frac{u}{R} \right)^2 \left( -f(u) d t^2 + \delta_{ij} d x^i d x^j +
d x_4^2 \right) + \left ( \frac{R}{u} \right)^2 \frac{du^2}{f(u)}, \\
F_6&=&Q_c \left(\frac{u}{R}\right)^4dx_0\wedge dx_1\wedge dx_2\wedge dx_3
                \wedge du\wedge dx_4 ,\\
e^\phi& =& \frac{2\sqrt{2}}{\sqrt{3}Q_c}, \qquad  R^2=\frac{15}{2}, \qquad
          f(u)=1-\left(\frac{u_T}{u}\right)^5 ,
\end{eqnarray}
where $i=\{1,2,3\}$, ${F_6}$ is the 6-form field strength of 5-form field coupled with
the color branes, ${Q_c}$ corresponds to the number of color branes ${N_c}$ and
${\phi}$ is the dilaton vacuum. The space spanned by $t$ and $u$ has the topology of
cigar with the minimum $u_T$ at the tip.
To avoid such a conical singularity, $t$ has to be periodic
\begin{equation}
t\sim t+\beta \sim t +\frac{4\pi R^2}{5u_T},
  \qquad \beta=\frac{4\pi R^2}{5u_T}.
\label{eq:temperature}
\end{equation}
Then we can take $T=1/\beta$ as the temperature of the black hole background geometry
and interpret it as the temperature of the dual boundary field theory. The corresponding background geometry of confined phase can be found in \cite{Mazu}.

We now briefly describe some physical aspects of this non-critical holographic QCD model. The main conclusions of this paragraph can be found in \cite{Mazu,Mintakevich}. As mentioned in section~\ref{introduction}, there are two kinds of phase transition for this model---one is the confinement/deconfinement phase transition and the other is the chiral phase transition, which is same as the critical Sakai-Sugimoto model. The former transition is realized as two different background geometries according to Witten \cite{Witten1}. Referring to the second one, it is geometrically realized as a thermal phase transition on the dual gravity side, marked by the shape of embedding profile of flavor D4/$\overline{\text{D4}}$ branes. This can be justified by comparison of on-shell actions (which are related to the free energies for the two phases respectively) for different flavor embedding profiles. The parallel D4/$\overline{\text{D4}}$ branes stands for chiral symmetric phase, while the U-shape flavor profile says that chiral symmetry is breaking. This identification can also be confirmed by studying the meson spectrum under these two different embedding profiles. As in the critical Sakai-Sugimoto model, one can find that the meson spectrum is discrete and gapped when the embedding profile is of U-shape; while for the parallel flavor embedding profile, there is no discrete meson spectrum at all. Generally, there is no parallel flavor profile in the confined phase due to U-shape topology of the confined background geometry in the $x_4-u$ plane. When one increases the temperature to the deconfined phase, there can be two kinds of flavor embedding profiles---parallel versus U-shaped. Therefore, the high temperature corresponds to the chiral symmetry restoring phase while the low temperature to the chiral broken phase. One striking feature of non-critical holographic QCD model is that the curvature of supergravity background is of order one and even cannot be suppressed by taking large $N_c$ limit. This is necessary to avoid an unwanted gap between low and high spin hadron spectra. However, unlike the case of the critical supergravity there is a problem with the possible stringy corrections to this non-critical supergravity. One may concern that the large stringy correction to the non-critical supergravity may affect every calculation on the gravity side, however the results in \cite{Kuperstein1} seem to indicate that this non-critical supergravity approximaton to the low energy behavior of non-critical superstring is meaningful. Another good feature of this non-critical model is that its dilaton vacuum is constant and does not need to be elevated to M-theory setup in the ultraviolet region as in the critical Sakai-Sugimoto model.

For later convenience of extraction of the light quark diffusion constant, we now turn to the calculation of the charge susceptibility in the chiral symmetric phase. To achieve this, we turn on the chemical potential\footnote{Chiral phase transition of this model at finite chemical potential has been studied in \cite{Cui1}). Note that conventions used in our study have some differences from \cite{Cui1}.} and its corresponding charge density which can be simulated by the time component of U(1) (in our study we will not consider the non-Abelian part of U($N_f$) flavor gauge group) gauge field on the flavor worldvolume. So in the following computation, we just assume $A_0(u)\neq 0$ and set other components of the flavor gauge field to be zero. The total effective action for D-brane includes the Dirac-Born-Infeld (DBI) part and the Chern-Simons part, but we here omit the Chern-Simons term due to the reason given in \cite{Mazu}\footnote{In actual fact, the Chern-Simons term only has some effect on the embedding profile of the flavor probe branes in the chiral broken phase, which will not be considered in this work. However, the qualitative feature of the profile in the chiral broken phase is still unchanged when adding the Chern-Simons term, i.e., it is still of U-shape. While in the chiral symmetric phase, we will set $x_{4}^{\prime}(u)=0$ and then contribution to the total D-brane action from the Chern-Simons term is exactly zero. What's more, the fluctuation around $x^{\prime}_{4}(u)=0$ just induces a linear term in the scalar action, and therefore do not contribute to the equation of motion. Thus, we can conclude that omitting the Chern-Simons term in this work will not change our main results.}: if Chern-Simons term were included in the total action, free energies difference between chiral symmetric and chiral breaking phases is divergent. Once omitting the Chern-Simons term, the action for flavor probe branes is:
\begin{equation}
S_{DBI}=-T_4 N_f \int d^4x du e^{-\phi} \sqrt{-\det{(g_{ind} +2\pi l^2_s F)}}\, ,
\end{equation}
where $g_{ind}$ is the induced metric on the flavor worldvolume, $T_4$ is the tension of flavor D4-brane, $l_s$ is the string length and the only nonzero component of field strength is $F_{0u}=F_{u0}$. The induced metric is given by
\begin{equation}
ds_{ind}^2=\left( \frac{u}{R} \right)^2 \left( -f(u) d t^2 + \delta_{ij} d x^i d x^j +
{x_4^{\prime}}^2 du^2 \right) + \left ( \frac{R}{u} \right)^2 \frac{du^2}{f(u)},
\end{equation}
where the prime means derivative with respect to the radial coordinate $u$ and $x_4(u)$ is the flavor brane embedding profile in this $AdS_6$ black hole background. Since we will focus on the chiral symmetric phase, we can set $x_4^{\prime}$ to be zero and the action can be greatly simplified
\begin{eqnarray}
S_{DBI}&=&V_4 \tilde{S}_{DBI} ,\\
\tilde{S}_{DBI}&=&-\frac{T_4 N_f}{e^\phi} \int du \left(\frac{u}{R}\right)^3
\sqrt{1-{\left(2 \pi l^2_s A_0^{\prime} \right)}^2}\, .
\end{eqnarray}
Here $V_4$ is the volume of the four dimensional Minkowski spacetime where the boundary field theory lives, and $\tilde{S}_{DBI}$ is the action density. Now we can get the equation of motion for $A_0(u)$:
\begin{eqnarray}
&& \frac{T_4 N_f}{e^\phi} \left(\frac{u}{R}\right)^3 \frac{\left(2\pi l^2_s\right)^2
A^{\prime}_0} {\sqrt{1-{\left(2 \pi l^2_s A_0^{\prime} \right)}^2}}=d \, ,
\label{eq:eomA}\\
&& \frac{T_4 N_f}{e^\phi}2\pi l^2_s \tilde{d}=d \, ,
\label{eq:d}
\end{eqnarray}
where we have chosen the integration constant as the physical charge density $d$ which is clear from holographic dictionary \cite{Karch2}, and the rescaled density $\tilde{d}$ is for convenience.

From above equations we obtain the relation between the chemical potential $\mu$ and the charge density as
\begin{equation}
2\pi l^2_s \mu=2\pi l^2_s\int^\infty_{u_T} A^\prime_0 du
=\int_{u_T}^\infty du \frac{\tilde{d}}
{\sqrt{\tilde{d}^2+\left(\frac{u}{R}\right)^6}} .
\label{eq:mu}
\end{equation}
From eqs.~(\ref{eq:d}) and (\ref{eq:mu}), we know that the physical charge density $d$ is proportional to the rescaled one $\tilde{d}$ and $\mu=0$ is equivalent to $d=0$. Therefore we can calculate the charge susceptibility $\Xi$ as
\begin{equation}
\Xi\equiv\left.\frac{\partial d}{\partial\mu}\right|_{\mu=0}=\left.\frac{\partial d}
{\partial\mu}\right|_{d=0}=\left.\left(\frac{\partial \mu}{\partial \tilde{d}}\right)^{-1}
\right|_{\tilde{d}=0} \frac{\partial\tilde{d}}{\partial d} . \label{eq:xi}
\end{equation}
The two parts on the right hand side of eq.~(\ref{eq:xi}) can be easily worked out from eqs.~(\ref{eq:mu}) and (\ref{eq:d}). Now we write down the final result for the charge susceptibility:
\begin{equation}
\Xi=\frac{2T_4 N_f}{e^\phi}\frac{\left(2\pi l^2_s \right)^2 u^2_T}{R^3}.
\end{equation}

We now change the notation conventions through the radial coordinate transformation $r=u_T/u$. Then the background geometry changes to the following form
\begin{eqnarray}
ds^2&=&\left( \frac{u_T}{R} \right)^2 \frac{1}{r^2}\left( -f(r) d t^2
    + \delta_{ij} d x^i d x^j +
d x_4^2 \right) + \left ( \frac{R}{r} \right)^2 \frac{dr^2}{f(r)}, \\
e^\phi& =& \frac{2\sqrt{2}}{\sqrt{3}Q_c}, \qquad  R^2=\frac{15}{2},
 \qquad  f(r)=1-r^5 ,
\end{eqnarray}
and the induced metric on the flavor worldvolume in the chiral symmetric phase is
\begin{equation}
ds_{ind}^2=\left( \frac{u_T}{R} \right)^2 \frac{1}{r^2}\left( -f(r) d t^2
+ \delta_{ij} d x^i d x^j
\right) + \left ( \frac{R}{r} \right)^2 \frac{dr^2}{f(r)} .
\end{equation}
Under this coordinate transformation, the horizon $u=u_T$ changes to $r=1$ and the boundary $u=\infty$ to $r=0$. This transformation will facilitate later analytical studies.

\subsection{Equations of motion for vector and scalar fluctuations}
\label{section:eom}
We now extract equations of motion for the vector and scalar fluctuations which will be the key ingredients for later calculations. Here we just focus on the vector U(1) part of the U($N_f$)$\times$ U($N_f$) gauge group for the $N_f$ flavor brane-antibrane pairs. Then the DBI action can be expanded to quadratic order in the field fluctuations. It is well-known that the vector and scalar modes do not couple to each other when the vector background is not turned on just as in this paper, and then we can separately analyze their fluctuations.

First, we consider the U(1) gauge field fluctuation and expand the DBI action to the quadratic order in the gauge field fluctuation:
\begin{eqnarray}
S_{\text{gauge}}&=&-T_4 N_f \int d^4x dr e^{-\phi}
\sqrt{-\det{\left(g_{ind}+2\pi l_s^2 F\right)}}\nonumber\\
&=&-\frac{\left(2\pi l^2_s\right)^2}{4} \frac{T_4 N_f}{e^\phi}
\int d^4x dr \sqrt{-\det{g_{ind}}} F_{\alpha\beta}F^{\alpha\beta} ,\label{1}
\end{eqnarray}
where $\alpha$ and $\beta$ denote $t,x,y,z$ and $r$. We will choose the radial gauge, i.e., $A_r=0$. Indices $\alpha$ and $\beta$ are contracted using the induced metric $g_{ind}$. The equation of motion for the gauge filed fluctuation is as follows,
\begin{equation}
\partial_\alpha \left(\sqrt{-\det{g_{ind}}} F^{\alpha\beta}\right)=0 .
\label{eq:eom}
\end{equation}
Inserting eq.~(\ref{eq:eom}) into eq.~(\ref{1}) results in the on-shell action for the gauge field part which is crucial in calculating the retarded Green's functions:
\begin{eqnarray}
S_{\text{gauge}}&=&-\frac{\left(2\pi l^2_s\right)^2}{2}
\frac{T_4 N_f}{e^\phi} \int d^4x dr
\partial_\alpha \left[\sqrt{-\det{g_{ind}}} A_\beta F^{\alpha\beta}\right]\nonumber\\
&=&-\frac{\left(2\pi l^2_s\right)^2}{2} \frac{T_4 N_f}{e^\phi} \int d^4x dr
\left\{\partial_r \left[\sqrt{-\det{g_{ind}}} A_\mu F^{r\nu}\right]\right.
\nonumber\\
&& \left. +\sqrt{-\det{g_{ind}}}
\partial_\mu\left[A_\nu F^{\nu\mu}\right]\right\}\label{eq:on-shell} ,
\end{eqnarray}
where $\mu$ and $\nu$ go from $t$ to $z$.

Now we perform Fourier transformation for the gauge potential $A_\mu$. Without loss of generality, we will choose the spatial momentum to be along the $x$ direction
\begin{eqnarray}
&& A_\mu\left(x_\mu,r\right)=\int\frac{d^4k}{(2\pi)^4} e^{ikx}
A_\mu\left(k_\mu,r\right) , \label{eq:A}\\
&& k_\mu=\left(-\omega,q,0,0\right) .
\label{eq:k}
\end{eqnarray}
Inserting eqs.~(\ref{eq:A}) and (\ref{eq:k}) into eq.~(\ref{eq:on-shell}), we get the final result for the on-shell action of the gauge field fluctuation as
\begin{eqnarray}
S_{\text{gauge}}&=&-\frac{\left(2\pi l^2_s\right)^2}{2} \frac{T_4 N_f}{e^\phi}
\frac{u^2_T}{R^3}\int \frac{d^4k}{(2\pi)^4}\frac{1}{r}
\left[f(r)A_i(-k,r)\partial_r A_i(k,r) \right.  \nonumber\\
&&~~~~~~~~~~~~~~~~~~~~~~~~~~~~~~~~~~~~~~~~~
\left. \left. -A_t(-k,r)
\partial_r A_t(k,r)\right]\right|^{r=1}_{r=0},
\end{eqnarray}
where $i$ goes from $x$ to $z$. Choosing $\beta$ = $r$, $x$, $y$ in eq.~(\ref{eq:eom}), we obtain respectively
\begin{eqnarray}
&& \omega A^\prime_t + q f(r) A^\prime_x=0 ,\\
&& \partial_r \left(\sqrt{-\det{g_{ind}}} g^{rr} g^{xx}A^\prime_x\right)
   -\sqrt{-\det{g_{ind}}}
   g^{tt} g^{xx}\omega\left(\omega A_x +q A_t\right)=0 ,\\
&& \partial_r \left(\sqrt{-\det{g_{ind}}} g^{rr}
 g^{yy}A^\prime_y\right)-\sqrt{-\det{g_{ind}}}
g^{yy}\left(\omega^2 g^{tt} +q^2 g^{yy}\right) A_y=0 .
\label{eq:Ay}
\end{eqnarray}
The $A_z$ component obeys similar equation as $A_y$. We clearly see that equations of motion for $A_t$ and $A_x$ couple together. Following the approach of \cite{Kovtun2}, we now define $E_L\equiv\omega A_x + q A_t$ and $E_T\equiv E_{y,z}=\omega A_{y,z}$ as the longitudinal and transversal electric fields respectively, then they obey the following decoupled equations of motion:
\begin{eqnarray}
&& E^{\prime\prime}_T+\left(\frac{f^\prime(r)}{f(r)}-\frac{1}{r}\right) E^\prime_T+
\frac{\tilde{\omega}^2-\tilde{q}^2 f(r)}{f^2(r)}E_T=0, \label{eq:T} \\
&& E^{\prime\prime}_L-\left(\frac{\tilde{\omega}^2
 f^\prime(r)}{f(r)\left(\tilde{q}^2 f(r)-
\tilde{\omega}^2\right)}+\frac{1}{r}\right) E^\prime_L
 +\frac{\tilde{\omega}^2-\tilde{q}^2
f(r)}{f^2(r)}E_L=0 . \label{eq:L}
\end{eqnarray}
Now we can also write the on-shell action of the gauge field fluctuation in terms of the longitudinal and transversal electric fields $E_L$, $E_T$,
\begin{eqnarray}
S_{\text{gauge}}&=&-\frac{\left(2\pi l^2_s\right)^2}{2}
   \frac{T_4 N_f}{e^\phi} R \int\frac{d^4k}
{(2\pi)^4}\frac{1}{r}\left[\frac{f(r)}{\tilde{q}^2 f(r)
  -\tilde{\omega}^2}E_L\partial_rE_L \right.  \nonumber\\
&&\left. \left.~~~~~~~~~~~~~~~~~~~~~~~~~~~~~~~~~~~~~
-\frac{f(r)}{\tilde{\omega}^2}\left(E_y\partial_rE_y+
E_z\partial_rE_z\right)\right]\right|^{r=1}_{r=0} .
\label{eq:v-on-shell}
\end{eqnarray}
In above equations, we have rescaled the frequency and spatial momentum to make them dimensionless
\begin{equation}
\tilde{\omega}\equiv\frac{R^2}{u_T}\omega=\frac{\omega}{0.8\pi T}, \quad
\tilde{q}\equiv\frac{R^2}{u_T} q=\frac{q}{0.8\pi T},
\end{equation}
where $T$ is the temperature defined below eq.~(\ref{eq:temperature}). In the following sections, we will mainly use these dimensionless quantities to carry out our calculations.

Following the same procedure, we can get similar results for the scalar mode fluctuation, i.e., the fluctuation around the parallel embedding profile $x^\prime_4=0$. For conciseness, we here merely list the main equations and omit the cumbersome details. The fluctuation around $x^\prime_4=0$ can be described as
\begin{eqnarray}
ds^2&=&\left( \frac{u_T}{R} \right)^2 \frac{1}{r^2}\left( -f(r) d t^2 +
\delta_{ij} d x^i d x^j \right) + \left ( \frac{R}{r} \right)^2
\frac{dr^2}{f(r)}+
\left( \frac{u_T}{R} \right)^2 \frac{1}{r^2}
\partial_\alpha\varphi\partial_\beta\varphi
dx^\alpha dx^\beta\nonumber\\
&=&ds^2(g_{ind})+ds^2(g_2) .
\end{eqnarray}
Expanding the action to quadratic order in the scalar fluctuation $\varphi$ gives
\begin{eqnarray}
S_{\text{scalar}}&=&-T_4 N_f\int d^4x dr e^{-\phi}
     \sqrt{-\det{\left(g_{ind}+g_2\right)}}\nonumber\\
&=&-\frac{T_4 N_f}{2 e^\phi}\left(\frac{u_T}{R}\right)^2\int d^4x
 dr\sqrt{-\det{g_{ind}}}
\frac{1}{r^2}g^{\alpha\beta}\partial_\alpha\varphi\partial_\beta\varphi .
\end{eqnarray}
We can directly write down the equation of motion for the scalar mode
\begin{equation}
\partial_\alpha\left(\sqrt{-\det{g_{ind}}}g^{\alpha\beta}
\partial_\beta\varphi\right)=0.
\end{equation}
After Fourier transformation of the scalar mode $\varphi$
\begin{eqnarray}
&&\varphi\left(x_\mu,r\right)=\int\frac{d^4k}{(2\pi)^4} e^{ikx}\varphi
\left(k_\mu,r\right),\\
&&k_\mu=\left(-\omega,q,0,0\right),
\end{eqnarray}
the on-shell action of the scalar part and the equation of motion for $\varphi$ take the following forms:
\begin{eqnarray}
&& S_{\text{scalar}}=-\frac{T_4 N_f}{2 e^\phi}\frac{u^6_T}{R^7}
\int \frac{d^4k}{(2\pi)^4}\left.\frac{f(r)}{r^5}\varphi(-k,r)
\partial_r\varphi(k,r)\right|_{r=0}^{r=1},
\label{eq:scalar}\\
&& \varphi^{\prime\prime}+\left(\frac{f^\prime(r)}{f(r)}-\frac{5}{r}\right)
\varphi^\prime+
\frac{\tilde{\omega}^2-\tilde{q}^2 f(r)}{f^2(r)}\varphi=0.
\label{eq:seom}
\end{eqnarray}

Before closing this section, we give a brief remark about the equations of motion and the on-shell actions. From eqs.~(\ref{eq:T}) and (\ref{eq:seom}) we can find that the transversal electric field $E_L$ and the scalar mode $\varphi$ have similar behaviors, and the only difference between them is in $1/r$ versus $5/r$ term in their equations which will affect their solutions near the boundary $r=0$. When we choose the spatial momentum to be zero, the equations obeyed by the longitudinal and transversal electric fields $E_L,E_T$ coincide. Now we turn to the on-shell actions given in eqs.~(\ref{eq:v-on-shell}) and (\ref{eq:scalar}). Once we have solved these equations of motion, we can insert these solutions into eqs.~(\ref{eq:v-on-shell}) and (\ref{eq:scalar}) and then we do second functional derivatives of the actions with respect to the boundary values of $E_L$, $E_T$ and $\varphi$. The results are just the retarded Green's functions if we merely retain the boundary value of the action. This is
proposed in \cite{Son1} using AdS/CFT technique to calculate the operator correlation function of the dual boundary thermal field theory. As pointed out in \cite{Son1}, we should only retain the boundary value of the action in order to satisfy some basic properties of the retarded Green's function. This does not mean that we just need to focus on the boundary behavior of these equations. The last key point under this proposition is that all the solutions to these equations of motion should satisfy incoming wave boundary conditions at the horizon, manifesting the unreflecting feature of black hole. This directly determines some important properties of the final results which will be explicit in the following sections.

\section{Hydrodynamic limit and light quark diffusion constant}
\label{sec:low}
\subsection{Retarded Green's function from gauge/gravity duality}
In this subsection, we first give a short summary of the basic prescription for calculating the Minkowskian retarded Green's function under gauge/gravity duality approach, and then we take the scalar operator as an example to show how to put this prescription into pratice. The main content of this subsection can be found in the original papers such as \cite{Son,Son1}, and we take it here just for self-consistency of our work. One important idea under the framework of gauge/gravity duality is the Operator/Field correspondence. Explicitly, this means that the operator on the dual field theory side is in one-one correspondence with the field in curved background geometry. Another striking feature of gauge/gravity duality is that this correspondence is a weak/strong duality and therefore we can turn strongly-coupled field theoretical problems into weakly-interacting gravitational ones. What's more, one can use the semi-classical approximation to deal with quantum field theory in curved spacetime under weak coupling limit of gravity, i.e., taking the supergravity geometry as a background. In conclusion, this duality tells us that the main physics for a specific system are now encoded in the classical equations of motion for the fields in curved spacetime.

The prescription of \cite{Son1} for computing the Minkowskian retarded Green's functions can be simply organized as the following two steps:

(1) Find a solution to the equation of motion for the mode fluctuation such as eq.~(\ref{scalar}) below with the following two properties: the first one is that the solution is set to be 1 at the boundary (i.e., one should impose Dirichlet condition for the fluctuation at the conformal boundary); the second one is that for timelike momenta, the solution should take an asymptotic expression corresponding to the incoming wave at the horizon while for a spacelike momenta, the solution should be regular at the horizon.

(2) The retarded Green's function is then given by a formula such as eq.~(\ref{two point function}) (roughly speaking, the second derivative of the on-shell action with respect to the boundary value of the corresponding mode fluctuation). In detail, this means that when one takes second order derivative of the on-shell action with respect to the boundary value of the mode fluctuation, only the contribution from the conformal boundary has to be taken. Surface terms from the horizon must be dropped. This part of metric affects the retarded Green's functions only through the boundary condition imposed for the bulk field fluctuation.

One important feature of this prescription, which has been pointed out in \cite{Son1}, is that the imaginary part of $\mathcal {F}(k,z)$ (see below) is independent of the radial coordinate. Therefore, the imaginary part of the retarded Green's function can be calculated at any convenient value of radial coordinate. The authors of \cite{Son1} have also checked their prescription by computing the retarded Green's functions in theories where they are known from other methods.

Now we turn to a scalar operator $\phi(x,z)$ as an example to elaborate detailed calculation of the Minkowskian retarded Green's function under gauge/gravity duality, which can be found in the original paper \cite{Son}. The action to quadratic order in scalar fluctuation takes the following form,
\begin{equation}
S=K \int d^4x\int_{z_B}^{z_H}dz\sqrt{-g}\left[g^{zz}\left(\partial_z\phi\right)^2+ g^{\mu\nu}\partial_{\mu}\phi\partial_{\nu}\phi+m^2 \phi^2\right],
\end{equation}
where $K$ is a normalization constant, $m$ is the mass of the scalar field, $z_B$ and $z_{H}$ are the locations of boundary and horizon respectively; $g^{zz}$ and $g^{\mu\nu}$ is the gravitational metric
\begin{equation}
ds^2=g_{zz}dz^2+g_{\mu\nu}dx^{\mu}dx^{\nu}.
\end{equation}
The linearized field equation for $\phi(x,z)$ is
\begin{equation}
\frac{1}{\sqrt{-g}}\partial_{z}\left(\sqrt{-g}g^{zz}\partial_{z}\phi\right) + g^{\mu\nu} \partial_{\mu}\partial_{\nu}\phi -m^2 \phi=0.
\end{equation}

After doing partial Fourier transformation along $x^{\mu}$ for $\phi(x,z)$
\begin{equation}
\phi(x,z)=\int\frac{d^4k}{\left(2\pi\right)^4}e^{ik\cdot x} f_{k} (z)\phi_{0}(k),
\end{equation}
the equation of motion for the scalar mode then takes the following form
\begin{equation}
\frac{1}{\sqrt{-g}}\partial_{z}\left(\sqrt{-g}g^{zz}\partial_{z}f_{k}\right) - \left(g^{\mu\nu} k_{\mu}k_{\nu} + m^2 \right)\phi=0.\label{scalar}
\end{equation}

Under above equation of motion, the on-shell action for the scalar fluctuation now reduces to the surface term
\begin{eqnarray}
&&S=\int\frac{d^4}{\left(2\pi\right)^4}\phi_0(-k)\mathcal{F}(k,z)\phi_0(k)|^{z=z_H}_ {z=z_B},\\
&&\mathcal{F}(k,z)=K \sqrt{-g}g^{zz}f_{-k}(z)\partial_{z}f_{k}(z),
\end{eqnarray}
here we have chosen the boundary value for $f_{k}(z)$ to be unit for convenience. Then the prescription for retarded Green's function is that we should just take the boundary value of on-shell action. The retarded Green's function is therefore proposed to be
\begin{equation}
G^{R}(k)=-2\mathcal{F}(k,z)|_{z=z_B}. \label{two point function}
\end{equation}
Though eq.~(\ref{two point function}) is just a proposal, its validity has been checked in the original papers and it plays an important role in many interesting problems.

The last important point in above calculation is that the linearized equation of motion for $\phi(k,z)$ should obey incoming wave boundary condition at the horizon $z=z_{H}$ due to fully-absorbing feature of black hole, which has been pointed out at the end of last section.

However, the classical equation of motion such as eq.~(\ref{scalar}) cannot be solved by analytical method in generic case. Therefore, many works have turned to numerical derivations of thermal retarded Green's function under gauge/gravity duality. In some limiting cases like hydrodynamic or high frequency limits, we have semi-analytical methods to solve these complicated second order differential equations. In detail, for hydrodynamic limit, we can have a double series expansion of solution in terms of dimensionless frequency $\tilde{\omega}$ and spatial momentum $\tilde{q}$; while for high frequency case, we can use WKB approximation to construct semi-classical wave functions for different modes. For WKB construction of the solution, we should carefully study these equations of motion near its regular singularities, the boundary and horizon, to get asymptotic solutions near these points and match these solutions with WKB solutions in the regular regions. This matching will finally determine the boundary solutions for different modes. In the remainder of this section, we will in detail solve equations of motion derived in section~\ref{section2} in hydrodynamic limit. We will present the detailed construction of WKB solutions in section~\ref{sec:high}.
\subsection{Analytical result in hydrodynamic limit}
Before detailed analysis, we need connect asymptotic solutions to eqs.~(\ref{eq:T}), (\ref{eq:L}) and (\ref{eq:seom}) near the boundary $r=0$ with the retarded Green's functions of dual mesonic operators. It is simple to show that near $r=0$ exponents of these equations are (0,2), (0,2) and (0,6) respectively. Then their general solutions can be written as
\begin{eqnarray}
E_T\left(k,r\right)&=&\mathcal {A}_T\left(\tilde{\omega},\tilde{q}\right)
\left(1+\ldots\right)+\mathcal {B}_T\left(\tilde{\omega},\tilde{q}\right)
\left(r^2+\ldots\right),\label{boundary for ET}\\
E_L\left(k,r\right)&=&\mathcal {A}_L\left(\tilde{\omega},\tilde{q}\right)
\left(1+\ldots\right)+\mathcal {B}_L\left(\tilde{\omega},\tilde{q}\right)
\left(r^2+\ldots\right),\label{boundary for EL}\\
\varphi\left(k,r\right)&=&\mathcal {A}_\varphi\left(\tilde{\omega},
\tilde{q}\right)\left(1+\ldots\right)+\mathcal {B}_\varphi
\left(\tilde{\omega},\tilde{q}\right)\left(r^6+\ldots\right),
\label{boundary for phi}
\end{eqnarray}
where `$\ldots$' indicates higher powers of $r$. Precisely speaking, theses solutions are not the general ones because the difference of each pair of exponents is an integer. Inversely, the general solution should have logarithmic terms, which may induce contact terms in the retarded Green's function as pointed out in \cite{Son1}. Given that these terms are not important in our extraction of the quark diffusion constant, hereafter we will omit them and take above solutions as the general ones. Inserting these boundary solutions into on-shell actions given in eqs.~(\ref{eq:v-on-shell}) and (\ref{eq:scalar}), we can easily obtain the following connections for different modes (the proportional coefficients for different modes will be explicitly given later):
\begin{equation}
G^R\left(\tilde{\omega},\tilde{q}\right)
\sim\frac{\mathcal {B}\left(\tilde{\omega},
\tilde{q}\right)}{\mathcal {A}\left(\tilde{\omega},\tilde{q}\right)},
\end{equation}
where $G^R\left(\tilde{\omega},\tilde{q}\right)$ is the retarded Green's function for dual mesonic operator.

Hydrodynamic limit means $\omega\ll T$ and $q\ll T$, or equivalently $\tilde{\omega}\ll 1$ and $\tilde{q}\ll 1$. This fact allows us to follow \cite{Policastro1,Policastro} to do double series expansion for $E_T(k,r)$, $E_L(k,r)$ and $\varphi(k,r)$ near
$\tilde{\omega}=0$ and $\tilde{q}=0$, which has been intensively studied in the literature. Near the horizon $r=1$, exponents of differential eqs.~(\ref{eq:T}), (\ref{eq:L}) and (\ref{eq:scalar}) are $\pm i\tilde{\omega}/5$. We should choose incoming wave boundary conditions at the horizon for different modes to reflect the unreflecting characteristic of black hole. This fact means that solutions to eqs.~(\ref{eq:T}), (\ref{eq:L}) and (\ref{eq:scalar}) near the horizon should take the following forms
\begin{equation}
E_{T,L}\left(k,r\rightarrow 1\right) ~\text{and}~\varphi
\left(k,r\rightarrow 1\right)\sim f(r)^{-i\tilde{\omega}/5}.
\end{equation}
As a result, we can make this condition more explicitly by assuming that the general solutions to these equations are of the following forms
\begin{eqnarray}
E_T\left(k,r\right)&=&f(r)^{-i\tilde{\omega}/5} E_T(r),\\
E_L\left(k,r\right)&=&f(r)^{-i\tilde{\omega}/5} E_L(r),\\
\varphi\left(k,r\right)&=&f(r)^{-i\tilde{\omega}/5} \varphi(r) .
\label{expansion for varphi}
\end{eqnarray}
Reasonably, we will also impose regularity of $E_T(r)$, $E_L(r)$ and $\varphi(r)$ at the horizon $r=1$.

It is not difficult to show that these new variables $E_T(r)$, $E_L(r)$ and $\varphi(r)$ obey the following equations of motion:
\begin{eqnarray}
&&E^{\prime\prime}_T(r)+\left[\frac{2i\tilde{\omega}r^4}{f(r)}
+\frac{f^\prime(r)}{f(r)}-\frac{1}{r}\right]E^\prime_T(r)\nonumber\\
&&+\left[\frac{4i\tilde{\omega}r^3}{f(r)}
-\frac{\tilde{\omega}^2r^8-5i\tilde{\omega}r^8}
{f^2(r)}+\left(\frac{f^\prime(r)}{f(r)}-\frac{1}{r}\right)
\frac{i\tilde{\omega}r^4}{f(r)}+\frac{\tilde{\omega}^2
-\tilde{q}^2f(r)}{f^2(r)}\right]E_T(r)=0, \label{eq:Tmode}\\
&&E^{\prime\prime}_L(r)+\left[\frac{2i\tilde{\omega}r^4}{f(r)}
-\frac{\tilde{\omega}^2f^\prime(r)}
{f(r)\left(\tilde{q}^2f(r)-\tilde{\omega}^2\right)}
-\frac{1}{r}\right]E^\prime_L(r)\nonumber\\
&&+\left[\frac{3i\tilde{\omega}r^3}{f(r)}
-\frac{\tilde{\omega}^2r^8-5i\tilde{\omega}r^8}{f^2(r)}
+\frac{i\tilde{\omega}^3r^4f^\prime(r)}
{f^2(r)\left(\tilde{q}^2f(r)-\tilde{\omega}^2\right)}
+\frac{\tilde{\omega}^2-\tilde{q}^2f(r)}{f^2(r)}
\right]E_L(r)=0, \label{eq:Lmode}\\
&&\varphi^{\prime\prime}(r)+\left(\frac{2i\tilde{\omega}r^4}{f(r)}
+\frac{f^\prime(r)}{f(r)}-\frac{5}{r}\right)\varphi^\prime(r)\nonumber\\
&&+\left[\frac{4i\tilde{\omega}r^3}{f(r)}-\frac{\tilde{\omega}^2r^8
-5i\tilde{\omega}r^8}{f^2(r)}+\left(\frac{f^\prime(r)}{f(r)}
-\frac{5}{r}\right) \frac{i\tilde{\omega}r^4}{f(r)}
+\frac{\tilde{\omega}^2-\tilde{q}^2f(r)}{f^2(r)}\right]\varphi(r)=0 .
\label{eq:smode}
\end{eqnarray}
Before doing double series expansions for $E_T(r)$, $E_L(r)$ and $\varphi(r)$, we now briefly comment on these three equations. As has been remarked in section \ref{section:eom}, the transversal electric field mode $E_T$ obeys similar equations as the scalar mode $\varphi$ and the same thing also happens here. So in the following calculations, we can just focus on vector modes and their results can be simply generalized to the scalar case.

For $E_T(r)$, its double series expansion in terms of $\tilde{\omega}$ and $\tilde{q}$ is
\begin{equation}
E_T(r)=F_0(r)+\tilde{\omega}F_1(r)+\ldots
\label{eq:Texpansion}
\end{equation}
where `$\ldots$' represents higher orders in $\tilde{\omega}$ and $\tilde{q}$ which are not important in our calculations. Putting this expansion into eq.~(\ref{eq:Tmode}) and counting by order of $\tilde{\omega}$ give
\begin{eqnarray}
&& F^{\prime\prime}_0+\left(\frac{f^\prime(r)}{f(r)}-\frac{1}{r}\right)F^\prime_0=0,
\label{eq:F0}\\
&& F^{\prime\prime}_1+\left(\frac{f^\prime(r)}{f(r)}-\frac{1}{r}\right)F^\prime_1+
\frac{2ir^4}{f(r)}F^\prime_0+\frac{3ir^3}{f(r)}F_0=0 .
\label{eq:F1}
\end{eqnarray}
From eq.~(\ref{eq:F0}) we can get $F_0=C$. In obtaining this final result, we have imposed regularity condition at the horizon for $F_0$. Along this procedure, it is not hard to work out the solution for eq.~(\ref{eq:F1}):
\small
\begin{eqnarray}
F_1(r)&=&\frac{iC}{20}\left\{4\ln\frac{1-r^5}{1-r}-(\sqrt5-1)\ln\left[r^2+\frac{1}
{2}(1+\sqrt5)r+1\right]\right.\nonumber\\
&&+(\sqrt5+1)\ln\left[r^2-\frac{1}{2}(-1+\sqrt5)r+1\right]+2\sqrt{10-2\sqrt5}
\arctan\left(\frac{4r-\sqrt5+1}{\sqrt{10+2\sqrt5}}\right)\nonumber\\
&&-2\sqrt{10+2\sqrt5}\arctan\left(\frac{4r+\sqrt5+1}{\sqrt{10-2\sqrt5}}\right)-
\ln3125-\sqrt5\ln(5-\sqrt5)+\sqrt5\ln(5+\sqrt5)\nonumber\\
&&\left.+2\sqrt{10-2\sqrt5}\arctan\left(\frac{5-\sqrt5}{\sqrt{10+2\sqrt5}}\right)-
2\sqrt{10+2\sqrt5}\arctan\left(\frac{5+\sqrt5}{\sqrt{10-2\sqrt5}}\right)\right\} .
\label{solutiontoF1}
\end{eqnarray}
\normalsize
To get the final form of this solution, regularity together with vanishing conditions at the horizon for $F_1$ have been imposed. Now we proceed by expanding the solution for $E_T$ to $\mathcal {O}(r^2)$
\small
\begin{eqnarray}
E_T(k,r)&=&C\left\{\left[1+\frac{i}{20}\left(-2\sqrt{1-\frac{2}{\sqrt5}}\pi+
2\sqrt5\coth^{-1}(\sqrt5)-5\ln5\right)\tilde{\omega}\right]+\frac{1}{2}i
\tilde{\omega}r^2\right\}\nonumber\\
&\simeq &C\left(1+\frac{1}{2}i\tilde{\omega}r^2\right).
\end{eqnarray}
\normalsize
Then it is straightforward to write down the retarded Green's function for the transversal electric field mode $E_T$ in the hydrodynamic limit
\begin{eqnarray}
G^R_{TT}\left(\tilde{\omega},\tilde{q}\right)&=&-2\times\frac{\left(2\pi
l^2_s\right)^2}{2} \frac{T_4 N_f}{e^\phi}\frac{R}{\tilde{\omega}^2}
\frac{2\mathcal {B}_T\left(\tilde{\omega},\tilde{q}\right)}{\mathcal {A}_T
\left(\tilde{\omega},\tilde{q}\right)}\\
&=&-\frac{\left(2\pi l^2_s\right)^2 T_4 N_f R}{e^\phi}\frac{i}{\tilde{\omega}}\, .
\end{eqnarray}

We now give similar analysis for the longitudinal electric field mode $E_L$, which will be much more complicated. Following the original proposition in \cite{Policastro1,Policastro}, we should perform double series expansion for $E_L(r)$ as follows:
\begin{equation}
E_L(r)=G_0(r)+\tilde{\omega}G_1(r)+\frac{\tilde{q}^2}{\tilde{\omega}}G_2(r)+\ldots
\end{equation}
As in the transversal electric field mode $E_T$ case, here `$\ldots$' represents higher order terms in $\tilde{\omega}$ and $\tilde{q}$. We can immediately write down the equations of motion obeyed by $G_0(r)$, $G_1(r)$ and $G_2(r)$, but here we slightly alter our procedure, which will avoid lengthy differential equations. Equations obeyed by $G_0(r)$ are
\begin{eqnarray}
G^{\prime\prime}_0+\left(\frac{f^\prime(r)}{f(r)}-\frac{1}{r}\right)G^\prime_0=0 ,\\
G^{\prime\prime}_0-\frac{1}{r}G^\prime_0=0 .
\end{eqnarray}
Then solution to $G_0(r)$ is very simple, given by
\begin{equation}
G_0(r)=C^\prime .
\end{equation}
Equations for $G_1(r)$ and $G_2(r)$ are given by
\begin{eqnarray}
&&G^{\prime\prime}_1+\left(\frac{f^\prime(r)}{f(r)}-\frac{1}{r}\right)G^\prime_1+
\frac{3ir^3}{f(r)}C^\prime=0, \label{eq:G1}\\
&&G^{\prime\prime}_2+\left(\frac{f^\prime(r)}{f(r)}-\frac{1}{r}\right)G^\prime_2+
f^\prime(r)G^\prime_1-\frac{5ir^8}{f(r)}C^\prime=0 .
\label{eq:G2}
\end{eqnarray}
Fortunately, eq.~(\ref{eq:G1}) is exactly eq.~(\ref{eq:F1}) once we take $F_0=C^\prime$ in the latter equation. Then solution to eq.~(\ref{eq:G1}) is just eq.~(\ref{solutiontoF1}) in which we should substitute $C^\prime$ for $C$, i.e.,
\begin{equation}
G_1(r)=F_1(r)|_{C=C^\prime} .
\end{equation}
Putting the solution for $G_1(r)$ into eq.~(\ref{eq:G2}) and imposing regularity and vanishing conditions at the horizon lead to the following solution to $G_2(r)$:
\begin{equation}
G_2(r)=\frac{iC^\prime}{2}\left(1-r^2\right) .
\end{equation}
Just like the transversal electric field mode $E_T$ case, we can now expand $E_L(k,r)$ to $\mathcal {O}(r^2)$
\begin{eqnarray}
E_L(k,r)&=&C^\prime\left\{\left[1+\frac{i}{20}\left(-2\sqrt{1-\frac{2}{\sqrt5}}
\pi+2\sqrt5\coth^{-1}(\sqrt5) -5\ln5\right)\tilde{\omega}
+\frac{i}{2}\frac{\tilde{q}^2}
{\tilde{\omega}}\right] \right. \nonumber \\
&& \left. ~~~~~~~~~~~~~~~~~~~~~~~~~~~~~~~~~~~~~~~~~~~~~~~~~~~~~~~~~~~~~~~~
+\frac{1}{2}i\left(\tilde{\omega}-\frac{\tilde{q}^2}
{\tilde{\omega}}\right)r^2\right\}\nonumber\\
&\simeq &C^\prime\left[\left(1+\frac{i}{2}\frac{\tilde{q}^2}{\tilde{\omega}}\right)+
\frac{1}{2}i\left(\tilde{\omega}
-\frac{\tilde{\omega}^2}{\tilde{q}}\right)r^2\right],
\label{EL}
\end{eqnarray}
where we follow the discussions in \cite{Kovtun2} to omit some unimportant terms of $\tilde{\omega}$. The retarded Green's function for the longitudinal electric field mode $E_L$ can be straightforwardly obtained
\begin{eqnarray}
G^R_{LL}\left(\tilde{\omega},\tilde{q}\right)&=&-2\times\frac{\left(2\pi l^2_s\right)^2}{2}
\frac{T_4 N_f}{e^\phi}\frac{R}{\tilde{\omega}^2-\tilde{q}^2}\frac{2\mathcal {B}_L
\left(\tilde{\omega},
\tilde{q}\right)}{\mathcal {A}_L\left(\tilde{\omega},\tilde{q}\right)}\nonumber\\
&=&-\frac{\left(2\pi l^2_s\right)^2 T_4 N_f R}{e^\phi}\frac{i}{\tilde{\omega}
+\frac{i}{2}\tilde{q}^2} .
\end{eqnarray}

Note that the above retarded Green's functions are two point functions of gauge invariant $E_T$ and $E_L$ defined in section \ref{section:eom}. It is necessary to represent them in terms of the original gauge field $A_\mu$. This can be accomplished by the following transformation rules
\begin{eqnarray}
&&G^R_{tt}=q^2 G^R_{LL} ,\\
&&G^R_{xx}=\omega^2 G^R_{LL} ,\\
&&G^R_{tx}=G^R_{tx}=q\omega G^R_{LL} ,\\
&&G^R_{yy}=G^R_{zz}=\omega^2 G^R_{TT} ,
\end{eqnarray}
where the quantities like $G^R_{tt}$ and $G^R_{xx}$ are exactly correlation functions of the vector mesonic operators dual to the flavor gauge field $A_\mu$, which can be recognized as propagator of the vector meson.

We now explicitly write down all nonzero retarded Green's functions in terms of dimensional variables $\omega$ and $q$, and we have also revealed their dependence on the temperature ~$T$
\begin{eqnarray}
&&G^R_{tt}\left(\omega,q\right)=-\mathcal {N}T\frac{iq^2}{\omega+iDq^2} ,\nonumber\\
&&G^R_{xx}\left(\omega,q\right)=-\mathcal {N}T\frac{i\omega^2}{\omega+iDq^2} ,\nonumber\\
&&G^R_{xt}\left(\omega,q\right)=G_{tx}\left(\omega,q\right)=-\mathcal {N}T
\frac{iq\omega}{\omega+iDq^2}, \nonumber\\
&&G^R_{yy}\left(\omega,q\right)=G_{zz}\left(\omega,q\right)
=-\mathcal {N}T i\omega .
\label{retarded green function for vector}
\end{eqnarray}
The constant $D$ appeared in the longitudinal component of the vector meson correlation function is given by
\begin{equation}
D=\frac{1}{1.6\pi T} ,
\end{equation}
which is the light quark diffusion constant and will be discussed in detail later. The common factor $\mathcal {N}$ is
\begin{eqnarray}
\mathcal {N}=\frac{4\pi\left(2\pi l^2_s\right)^2 T_4 N_f R}{5e^\phi}
=\frac{3\sqrt5 Q_c N_f}{20\pi g_s l_s} .
\end{eqnarray}
To get the final result for this common factor, we have substituted values for
$\text{D4}$ brane tension $T_4$, dilaton vacuum $e^\phi$ and curvature radius $R$.

We now give a brief remark on these retarded Green's functions for the vector mode. It is well-known that correlation function for global current operator can be decomposed into transverse and longitudinal parts in thermal relativistic quantum field theory, which has been explicitly revealed in our results. Moreover, in the long-time and long-wavelength limit, longitudinal and transversal parts of the correlation function have a universal characteristic dictated by hydrodynamics: the transversal part is nonsingular of frequency while the longitudinal part has a simple pole at $\omega=-iDq^2$, where $D$ is the charge diffusion constant. We have seen that this feature has also explicitly appeared in our results. This once again confirms the applicability of gauge/gravity duality technique to studies of strongly coupled thermal non-Abelian gauge theories. Hereafter we will also refer to longitudinal electric field mode $E_L$ as the diffusive channel. We will in section~\ref{light} state the extractions of the light quark diffusion constant in detail.

Having completed calculations for the vector mode, we can generalize these results to the scalar mode case. The procedure and result are quite similar to that of the transversal electric field mode $E_T$. We first expand $\varphi(r)$ as for $E_T(r)$
\begin{equation}
\varphi(r)=H_0(r)+\tilde{\omega}H_1(r)+\ldots
\end{equation}
Then equations obeyed by $H_0(r)$ and $H_1(r)$ can be simply worked out:
\begin{eqnarray}
&&H^{\prime\prime}_0+\left(\frac{f^\prime(r)}{f(r)}-\frac{5}{r}\right)H^\prime_0=0,
\label{eq:H0}\\
&&H^{\prime\prime}_1+\left(\frac{f^\prime(r)}{f(r)}-\frac{5}{r}\right)H^\prime_1+
\frac{2ir^4}{f(r)}H^\prime_0+\frac{3ir^3}{f(r)}H_0=0.
\label{eq:H1}
\end{eqnarray}

These equations are very similar to those for the transversal electric field modes $E_T$ and the only difference lies in $1/r$ versus $5/r$ term, but their effect on the final solution is not important. We then directly write down the final solutions of these equations\footnote{We have imposed regularity condition at the horizon for $H_0(r)$ and regularity as well as vanishing conditions at the horizon for $H_1(r)$.}
\begin{equation}
H_0(r)=D^\prime\label{H0}
\end{equation}
where $D^\prime$ is a constant and
\begin{eqnarray}
H_1(r)&=&\frac{3iD^\prime}{100}\left(2\sqrt{25+2\sqrt5}\pi
+10\sqrt5 \coth^{-1}\sqrt5
-100+5\ln5\right) \nonumber\\
&&+\frac{3}{5}iD^\prime r^5-\frac{1}{2}iD^\prime r^6+\mathcal {O}(r^7).
\label{H1}
\end{eqnarray}

As explained in eq.~(\ref{boundary for phi}), we only need to get the solution for $\varphi(k,r)$ to $\mathcal {O}(r^6)$, which can be obtained by inserting eqs.~(\ref{eq:H0}) and (\ref{H1}) into eq.~(\ref{expansion for varphi})
\begin{eqnarray}
\varphi(k,r)&=&D^\prime\left\{\left[1+\frac{3i}{100}
\left(2\sqrt{25+2\sqrt5}\pi+10\sqrt5 \coth^{-1}\sqrt5-100+5\ln5\right)
\tilde{\omega} \right. \right.\nonumber\\
&&~~~~~~~~\left.\left. +\frac{4}{5}i\tilde{\omega}r^5\right]
-\frac{1}{2}i\tilde{\omega}r^6\right\} .\label{varphi(k,r)}
\end{eqnarray}
Comparing eq.~(\ref{varphi(k,r)}) with eq.~(\ref{boundary for phi}), it is clear
that
\begin{equation}
\mathcal {A}_\varphi\left(\tilde{\omega},\tilde{q}\right)= D^\prime,
\qquad\mathcal {B}_\varphi\left(\tilde{\omega},\tilde{q}\right)
=-\frac{iD^\prime}{2}\tilde{\omega} \, .
\end{equation}
Then we can write down result of correlation function for scalar mesonic operator
\begin{eqnarray}
G^R_{\varphi\varphi}\left(\tilde{\omega},\tilde{q}\right)=2\times\frac{T_4 N_f}
{2e^\phi}\frac{u^6_T}{R^7}\frac{6\mathcal {B}_\varphi
\left(\tilde{\omega},\tilde{q}\right)}
{\mathcal {A}_\varphi\left(\tilde{\omega},\tilde{q}\right)}
=-\mathcal {N}^\prime T^6 i\omega \, ,
\label{retarded green function for scalar}
\end{eqnarray}
and the factor $\mathcal {N}^\prime$ is
\begin{equation}
\mathcal {N}^\prime=\frac{3T_4 N_f R^5}{e^\phi}\left(\frac{4\pi}{5}\right)^6 \, .
\end{equation}
We see that the dependence of the scalar mode correlation function on temperature has a surprising $T^6$ scaling which is different from the case of vector mode, but its dependence on frequency is same as the transverse vector mode.

Before closing this subsection, we briefly list the mesonic spectral function. It is proportional to the imaginary part of retarded Green's function and defined as
\begin{equation}
\mathfrak{R}\left(\omega,\textbf{q}\right)\equiv-2\text{Im}~G^R\left(\omega,\textbf{q}
\right) \, . \label{spectral function}
\end{equation}
The explicit expressions for all nonzero components of mesonic spectral function are listed here
\begin{eqnarray}
&&\mathfrak{R}_{tt}\left(\omega,q\right)
=2\mathcal {N}T\frac{q^2\omega}{\omega^2+D^2q^4}\, ,\nonumber\\
&&\mathfrak{R}_{xt}\left(\omega,q\right)=\mathfrak{R}_{tx}\left(\omega,q\right)=
2\mathcal {N}T\frac{q\omega^2}{\omega^2+D^2q^4}\, , \nonumber\\
&&\mathfrak{R}_{xx}\left(\omega,q\right)=2\mathcal {N}T\frac{\omega^3}{\omega^2+D^2q^4}\, , \nonumber\\
&&\mathfrak{R}_{yy}\left(\omega,q\right)=\mathfrak{R}_{zz}\left(\omega,q\right)=2\mathcal {N}T\omega\, , \nonumber\\
&&\mathfrak{R}_{\varphi\varphi}\left(\omega,q\right)=2\mathcal {N}^\prime T^6\omega
 \, .
\end{eqnarray}

\subsection{Light quark diffusion constant}\label{light}
Now we can extract the light quark diffusion constant based on above results. In fact, transport coefficients for the stress tensor or conserved charge have been widely studied in the literature, but those studies mainly focus on the gluon sector, e.g., shear/bulk viscosity and $R$-current diffusion constant. The analysis in \cite{Myers} has elaborately studied the flavor charge transport in D3/D7 brane system. In the following, we will follow \cite{Myers} to use three different methods to extract this constant and we will find that our result is same as that of the D2-brane background studied in \cite{Kovtun}. If we take a careful look at the metric for D2-brane and the metric used here, we find that they have the same black factor $f(r)=1-r^5$. This indicates that this black factor may have a direct relationship with the diffusion constant. Actually, from analysis of equations of motion in previous subsection, we also found the crucial role of this black factor in determining final results.

\subsubsection{From the pole of diffusive channel}
This method has been briefly described below results of longitudinal mode correlation function. We now in detail elaborate physical picture as well as some key formulae behind this way. By this method, we need to study some general properties of current-current correlator purely from viewpoint of finite-temperature field theory. Then these properties should explicitly emerge from calculations on dual gravitational side. We then extract the diffusion constant for conserved current by matching of these properties on both side.

We here follow \cite{Kovtun2,Myers} to present the general Lorentz index structure of thermal retarded Green's functions of conserved current in relativistic quantum field theory. Due to the presence of nonzero temperature, the Poincar\'{e} symmetry of relativistic quantum field theory reduces to translation and rotation invariance. If the conserved current is denoted as $J_{\mu}(x)$, current-current correlator in coordinate space is defined as follows
\begin{equation}
G^{R}_{\mu\nu}\left(x-y\right)=-i\theta\left(x^0-y^0\right)\langle\left[J_{\mu}(x),J_{\nu}(y) \right]\rangle,
\end{equation}
where $\langle...\rangle$ symbols expectation value in a translation invariant state and $\theta(x^0-y^0)$ is the step function. Then it can be Fourier transformed into momentum space,
\begin{equation}
G^{R}_{\mu\nu}\left(x-y\right)=\int\frac{d^4k}{\left(2\pi\right)^4}e^{ik\cdot (x-y)}G^{R}_{\mu\nu}(k). \label{cc in k}
\end{equation}
Conservation of current $J_{\mu}(x)$ implies that current-current correlator defined in eq.~(\ref{cc in k}) should satisfy following Ward identity
\begin{equation}
k^{\mu}G^{R}_{\mu\nu}(k)=0.\label{ward}
\end{equation}

In vacuum, the temperature is zero and therefore conservation of current eq.~(\ref{ward}) demands that current-current correlator should take the following form
\begin{equation}
G_{\mu\nu}^{R}(k^2)=P_{\mu\nu}\Pi(k^2),\quad P_{\mu\nu}=\eta_{\mu\nu}-\frac{k_{\mu}k_{\nu}}{k^2}.\label{lorenz invariant}
\end{equation}
While in thermal relativistic quantum field theory, due to presence of rotation symmetry, it is then convenient to split the projector $P_{\mu\nu}$ into transverse and longitudinal parts,
\begin{eqnarray}
&P_{\mu\nu}=P_{\mu\nu}^{T}+P_{\mu\nu}^{L},\nonumber\\
&P_{00}^{T}=0,\quad P_{0i}^{T}=0,\quad P_{ij}^{T}=\delta_{ij}-\frac{k_{i}k_{j}}{\vec{k}^2},\\
&P_{\mu\nu}^{L}=P_{\mu\nu}-P_{\mu\nu}^{T}.\nonumber
\end{eqnarray}
One can easily check that $k^{\mu}P^{T}_{\mu\nu}=k^{\mu}P^{L}_{\mu\nu}=0$, which indicates that quantities composed of $P_{\mu\nu}^{T}$ and $P_{\mu\nu}^{L}$ can naturally satisfy Ward identity.

In thermal field theory with rotation symmetry, current-current correlation function is completely determined by two scalar functions
\begin{equation}
G^{R}_{\mu\nu}(k)=P^{T}_{\mu\nu}\Pi^{T}(k)+P^{L}_{\mu\nu}\Pi^{L}(k).
\end{equation}
When $\Pi^{T}=\Pi^{L}$, this formula is exactly reduced to the Lorenz-invariant form eq.~(\ref{lorenz invariant}). Actually, when the spatial momenta are chosen to be zero, then the transversal and longitudinal parts of the current-current correlator coincide and we also come to the result eq.~(\ref{lorenz invariant}). Without loss of generality we can take the spatial momentum $\vec{k}$ to be along $x_1$ direction, i.e., $k_{\mu}=(-\omega,q,0,0)$ as in this note. Then all nonzero components of current-current correlator are
\begin{eqnarray}
&G^{R}_{x_2x_2}(k)=G^{R}_{x_3x_3}(k)=\Pi^{T}(\omega,q),\quad
G^{R}_{tt}(k)=\frac{q^2}{\omega^2-q^2}\Pi^{L}(\omega,q),\nonumber\\ &G^{R}_{tx_1}(k)=G^{R}_{x_1t}(k)=\frac{-\omega q}{\omega^2-q^2}\Pi^{L}(\omega,q), \quad G^{R}_{x_1x_1}(k)=\frac{\omega^2}{\omega^2-q^2}\Pi^{L}(\omega,q).
\end{eqnarray}

For a system in stable thermodynamic equilibrium, the low frequency/momentum behavior of scalar functions $\Pi^{T}$ and $\Pi^{L}$ is universal and is described by effective hydrodynamic theory, see for example \cite{forster}. Under hydrodynamic approximation, the main conclusion is that $\Pi^{T}$ is nonsingular as a function of $\omega$ because it does not couple to charge density fluctuations, and correlator involved conserved charge density must exhibit a hydrodynamic singularity whose dispersion relation should satisfy $\omega(q)\rightarrow 0$ when $q\rightarrow 0$. As a result, the longitudinal part of current-current correlator $\Pi^{L}\left(\omega, q\right)$ has a simple pole at $\omega=-i D q^2$ and this equality can be recognized as the hydrodynamic dispersion relation, where $D$ is the charge diffusion constant associated with current $J_{\mu}(x)$.

When coming to our present calculations, one can imagine that a conserved current $J_{\mu}(x)$ (here, U(1) flavor current) couples to the flavor gauge field $A_{\mu}(x,r)$, encoded in action $\displaystyle\int d^4x A_{\mu}(x,r=0)J^{\mu}(x)$, i.e., the boundary value of flavor gauge field $A_{\mu}(x,r=0)$ is taken as the source for flavor current on boundary field theory. Then form gauge/gravity duality, flavor current-current correlator can be equivalently calculated as the retarded Green's function of $A_{\mu}(x,r)$ as shown in detail in previous subsection. Therefore, combining general properties and results for flavor gauge field two point function in previous subsection like eq.~(\ref{retarded green function for vector}), we can easily find the longitudinal part of current-current correlator has a pole at
\begin{equation}
\omega=-i D q^2,\quad D=\frac{1}{1.6\pi T},
\end{equation}
where $D$ is the flavor charge diffusion constant. Therefore, diffusion constant is just a byproduct of two point function of vector mesonic operator.
\subsubsection{Membrane paradigm}
This way of calculating transport coefficients was first proposed in \cite{Kovtun} and further developed in \cite{Starinets,Iqbal}. The main advantage of this approach is that explicit formulae for various transport coefficients can be expressed in terms of metric components. This property is the result of universality of membrane paradigm in studying hydrodynamic behaviors of strongly coupled thermal field theories through gauge/gravity duality. Here, we follow the conventions of \cite{Kovtun}. There, the authors considered small fluctuations of black brane background and formulae for different transport coefficients were obtained from holographically derived Fick's law. Those formulae are for gluon sector and we can extend them to probe brane sector, such as flavor charge transport coefficients. This can be accomplished by considering perturbation of flavor field living on probe brane. Suppose that flavor gauge field coupled with the conserved current has the action
\begin{eqnarray}
&& S\sim\int d^{p+2}x \sqrt{-\det{g}}\left(\frac{1}{g^2_{eff}}
  F^{\mu\nu}F_{\mu\nu}\right) ,\nonumber\\
&& d^{p+2}x=dt d^px dr,\qquad \mu=\left\{t, x^i, r\right\}
~~(i=1,\ldots, p) ,
\end{eqnarray}
where $g_{eff}$ is the effective coupling. The conserved charge diffusion constant has the universal formula \cite{Kovtun,Myers}
\begin{equation}
D=\left.\frac{\sqrt{-\det{\tilde{g}}}}{\tilde{g}_{xx}g^2_{eff}\sqrt{-\tilde{g}_{tt}
\tilde{g}_{rr}}}\right|_{r=r_T}\int^\infty_{r_T}dr \frac{-\tilde{g}_{tt}\tilde{g}_{rr}
g^2_{eff}}{\sqrt{-\det{\tilde{g}}}}\, ,\label{universal}
\end{equation}
where `$\sim$' indicates the induced metric on the worldvolume of probe brane and $r_T$ represents the apparent horizon in the induced metric. With this universal formula for transport coefficients, the light quark diffusion in this model can be simply obtained\footnote{In our calculations, we use the untransformed metric, i.e. the $u$ coordinate.}
\begin{eqnarray}
D=\left.\frac{\left(\frac{u}{R}\right)^3}{\left(\frac{u}{R}\right)^2
\sqrt{\left(\frac{u}{R}
\right)^2f(u)\left(\frac{R}{u}\right)^2\frac{1}{f(u)}}}\right|_
{u=u_T}\int^\infty_{u_T}\frac{1}{\left(\frac{u}{R}\right)^3}
=\frac{1}{1.6\pi T} \, .
\end{eqnarray}

\subsubsection{Green-Kubo formula}
Besides above two methods, one use also Green-Kubo relation to compute flavor charge diffusion constant. As is well-known from non-equilibrium statistical physics \cite{hosoya,lifshitz}, all kinetic coefficients can be expressed, through Green-Kubo relation, as the correlation functions of the corresponding currents. In more detail, this relation connects the flavor charge diffusion constant $D$, charge susceptibility $\Xi$ and zero-momentum limit of vector spectral function $\mathfrak{R}_{x_1 x_1}(\omega,\textbf{q})$ as follows \cite{hosoya,lifshitz}
\begin{equation}
D\Xi=\lim_{\omega\to 0}\frac{\mathfrak{R}_{x x}\left(\omega, \textbf{q}=0 \right)}{2\omega} .
\end{equation}
The charge susceptibility $\Xi$ has been computed in section \ref{setup} and the zero limit of vector spectral function can be extracted from the previous subsection, eq.~(\ref{retarded green function for vector}). Then we have
\begin{eqnarray}
D=\frac{1}{\Xi}\lim_{\omega\to 0}\frac{\mathfrak{R}_{xx}
   \left(\omega, \textbf{q}=0 \right)}{2\omega}
=\frac{2T_4 N_f}{e^\phi}\frac{\left(2\pi l^2_s \right)^2 u^2_T}{R^3}
  \mathcal {N}T
=\frac{1}{1.6\pi T}\, .
\end{eqnarray}

We see that three different ways give same results for flavor charge diffusion constant, which shows that gauge/gravity is a powerful tool in studying hydrodynamic aspects of strongly coupled thermal field theory. It can also be recognized as a check of correctness of our calculations. Before concluding this section, we would like to have a short comparison between non-critical holographic QCD model and critical Sakai-Sugimoto model on the main results in this section. Because related investigations in critical Sakai-Sugimoto model have not been completely presented in the literature, we have in previous sections mainly compared our results with that of D3/D7 brane system. But we have sufficient reasons to believe that similar results of mesonic operator correlation functions can be emergent for Sakai-Sugimoto model. Concentrating on flavor charge diffusion constant, it is found in \cite{Evans} that $D=\frac{1}{2 \pi T}$. Therefore, one lesson from these studies is that light quark diffusion constant is proportional to $T^{-1}$, in this sense the non-critical holographic QCD model brings no new features to us. But in section~\ref{section6} we will find one difference between these two models for heavy quark diffusion constant.
\section{High frequency: the WKB method}
\label{sec:high}
In this section we discuss the high frequency behavior for retarded Green's functions and spectral functions of mesonic operators. Some relevant studies for $\mathcal{N}=4$ super-Yang-Mills plasma can be found in \cite{Teaney}. The main idea under high frequency limit is that one can construct the WKB solutions to equations of motion for different modes once transformed to Schr\"{o}dinger type, and then matching solutions at different regimes to obtain the boundary behaviors of wave functions $\psi_{T,L,S}$. The direct conclusion under present study is that incoming wave boundary condition at the horizon can be extended to be imposed near the boundary $r=0$. Physically, this is not difficult to understand: in high frequency limit, the effective Schr\"{o}dinger potential is negative, i.e., there is no potential barrier to prohibit propagation of flavor fields in $AdS_6$ black hole background; moreover, in the absence of potential barrier, there is no method to generate reflected waves near the boundary. Hence, an incoming wave boundary condition at the horizon is still valid near the boundary. In the remainder of this section, we will demonstrate this by detailed analysis of above Schr\"{o}dinger type equations.

For the sake of convenience in analyzing these intricate differential equations, we first convert them into Schr\"{o}dinger types through following transformations
\begin{eqnarray}
&& E_T\left(k,r\right)=\sqrt{\frac{r}{1-r^5}}\psi_T\left(k,r\right),
     \label{ETtransformation}\\
&& E_L\left(k,r\right)=\sqrt{r\left(-\tilde{q}^2+\frac{\tilde{\omega}^2}{1-r^5}\right)}
   \psi_L\left(k,r\right), \\
&& \varphi\left(k,r\right)=\sqrt{\frac{r}{1-r^5}}\psi_S\left(k,r\right).
\end{eqnarray}
Then eqs.~(\ref{eq:T}), (\ref{eq:L}) and (\ref{eq:scalar}) can be expressed in terms
of $\psi_T\left(k,r\right)$, $\psi_L\left(k,r\right)$ and $\psi_S\left(k,r\right)$ as
\small
\begin{eqnarray}
&&\psi^{\prime\prime}_T\left(k,r\right)+\frac{-3+4r^2\left[9r^3-2r^8+\tilde{q}^2
\left(-1+r^5\right)+\tilde{\omega}^2\right]}{4r^2\left(1-r^5\right)^2}
\psi_T\left(k,r\right)=0,\\
&&\psi^{\prime\prime}_L\left(k,r\right)+\frac{1}{4}\left\{-\frac{75\tilde{q}^4r^8}
{\left[\tilde{q}^2(1-r^5)-\tilde{\omega}^2\right]^2}
+\frac{2\tilde{q}^2\left[-15r^3+40r^8-2\tilde{q}^2(1-r^5)+2\tilde{\omega}^2\right]}
{(1-r^5)\left[\tilde{q}^2(1-r^5)-\tilde{\omega}^2\right]}\right.\nonumber\\
&&\left.+\frac{-3+4r^2\left(9r^3-2r^8+\tilde{\omega}^2\right)}{r^2
\left(1-r^5\right)^2}\right\}
\psi_L\left(k,r\right)=0,\\
&&\psi^{\prime\prime}_S\left(k,r\right)+\frac{-35+4r^2\left[15r^3-\tilde{q}^2 (1-r^5)+
\tilde{\omega}^2\right]}{4r^2\left(1-r^5\right)^2}\psi_S\left(k,r\right)=0 .
\end{eqnarray}
\normalsize
Note that we do not take large spatial momentum limit so that we can fix $q$ and even set it to be zero when taking large $\omega$ limit. Then above three intricate Schr\"{o}dinger type equations further reduce to the following two simple ones when choosing $q=0$
\begin{eqnarray}
&& \psi^{\prime\prime}_{T,L}\left(k,r\right)+\frac{-3+4r^2\left(9r^3-2r^8
   +\tilde{\omega}^2\right)}
{4r^2\left(1-r^5\right)^2}\psi_{T,L}\left(k,r\right)=0, \label{Ehigh}\\
&& \psi^{\prime\prime}_S\left(k,r\right)+\frac{-35+4r^2
   \left(15r^3+\tilde{\omega}^2\right)}
{4r^2\left(1-r^5\right)^2}\psi_S\left(k,r\right)=0 .
\end{eqnarray}

These two equations both have only two singularities at $r=0$ and $r=1$ in the range of $\left[0,1\right]$. Away from these two points, we can directly write down the WKB approximate solutions to these equations. However, this kind of solutions is meaningless when going close to these singularities, where one should take a careful look at the asymptotic behaviors of the original equations. At last, one needs to match the asymptotic solutions near singularities with the WKB ones. In the following discussions, we merely state detailed calculations for vector mode, and directly write down final results for the scalar mode for brevity.

In high frequency limit, i.e., $\tilde{\omega}\gg 1$, eq.~(\ref{Ehigh}) can be further simplified when away from its singular points
\begin{equation}
\psi^{\prime\prime}_{T,L}\left(k,r\right)+\frac{\tilde{\omega}^2}{(1-r^5)^2}
\psi_{T,L}\left(k,r\right)=0 .
\end{equation}
Now we define the canonical momentum $p(r)$ and action $s(r)$
\begin{equation}
p(r)=\frac{\tilde{\omega}}{1-r^5},
\qquad s(r)=\int^r_0 p(r)dr=\int^r_0 dr \frac{\tilde{\omega}}{1-r^5}
\end{equation}
and the two WKB solutions are given by
\begin{eqnarray}
\psi_1 \sim \frac{1}{\sqrt{p(r)}}\cos\left(s(r)+\phi_1\right),\nonumber\\
\psi_2 \sim \frac{1}{\sqrt{p(r)}}\sin\left(s(r)+\phi_2\right),\label{WKB}
\end{eqnarray}
where the phases $\phi_1$ and $\phi_2$ will be determined by matching above WKB solutions with the asymptotic one near the singular point $r=0$.

Near $r=0$, eq.~(\ref{Ehigh}) reduces to
\begin{equation}
\psi^{\prime\prime}_{T,L}\left(k,r\right)+\left(-\frac{3}{4r^2}+\tilde{\omega}^2
\right)\psi_{T,L}\left(k,r\right)=0 .
\end{equation}
Its general solution is the linear combination of the first and second kind of Bessel functions
\begin{equation}
\psi(r\sim 0)=C(1)\sqrt r J_1\left(\tilde{\omega}r\right)+C(2)\sqrt r Y_1\left(
\tilde{\omega}r\right), \label{asy near r=0}
\end{equation}
where the integration constants $C(1)$ and $C(2)$ will be determined later.

Near $r=1$, eq.~(\ref{Ehigh}) reduces to
\begin{equation}
\psi^{\prime\prime}_{T,L}\left(k,r\right)+\frac{\frac{1}{4}+\frac{\tilde{\omega}^2}
{25}}{(1-r)^2}\psi_{T,L}\left(k,r\right)=0,
\end{equation}
whose general solution is much simpler than the previous one:
\begin{equation}
\psi(r\sim 1)=C(3)\left(1-r\right)^{\frac{1}{2}-\frac{i\tilde{\omega}}{5}}+C(4)
\left(1-r\right)^{\frac{1}{2}+\frac{i\tilde{\omega}}{5}} .
\end{equation}
From eq.~(\ref{ETtransformation}) and incoming wave boundary condition imposed at the horizon, we immediately obtain
\begin{equation}
C(4)=0\Longrightarrow\psi(r\sim 1)=C(3)
 \left(1-r\right)^{\frac{1}{2}-\frac{i\tilde{\omega}}{5}} .
\end{equation}

To match these solutions, we need asymptotic forms of $p(r)$ and $s(r)$ near $r=0$ and $r=1$ respectively:
\begin{eqnarray}
&&p(r)=\frac{\tilde{\omega}}{1-r^5}\approx \tilde{\omega}
   \quad (r\rightarrow 0),\nonumber\\
&&p(r)=\frac{\tilde{\omega}}{1-r^5}\approx
   \frac{\tilde{\omega}}{5(1-r)}\quad (r\rightarrow 1),\nonumber\\
&&s(r)=\int^r_0 dr \frac{\tilde{\omega}}{1-r^5}
    \approx\tilde{\omega}r \quad (r\rightarrow 0),\nonumber\\
&&s(r)=\int^r_0 dr \frac{\tilde{\omega}}{1-r^5}\approx\frac{1}{5}
   \left[-\tilde{\omega}
\ln(1-r)+\tilde{\omega}\ln{\text{constant}}\right]
 \quad (r\rightarrow 1) .\label{asy for p and s}
\end{eqnarray}
The constant in the last equation is very intricate, so we do not present its expression here. Using asymptotic series expansion of Bessel function at large $\tilde{\omega}r$:
\begin{eqnarray}
J_1\left(\tilde{\omega}r\right)\sim \sqrt{\frac{2}{\pi\tilde{\omega}r}}
  \cos\left(\tilde{\omega}r-\frac{3\pi}{4}\right),\nonumber\\
Y_1\left(\tilde{\omega}r\right)\sim \sqrt{\frac{2}{\pi\tilde{\omega}r}}
  \sin\left(\tilde{\omega}r-\frac{3\pi}{4}\right),
\end{eqnarray}
eq.~(\ref{asy near r=0}) has the following asymptotic behavior at high frequency limit
\begin{equation}
\psi(r\sim 0)\sim C(1)\sqrt{\frac{2}{\pi\tilde{\omega}}}
\cos\left(\tilde{\omega}r-
\frac{3\pi}{4}\right)+C(2)\sqrt{\frac{2}{\pi\tilde{\omega}}}\sin
\left(\tilde{\omega}r-\frac{3\pi}{4}\right)\label{near 0} .
\end{equation}

Now it is straightforward to perform matching of solutions underlying different regimes. With $r\rightarrow 0$ parts of eq.~(\ref{asy for p and s}), matching of
eq.~(\ref{asy near r=0}) with eq.~(\ref{WKB}) gives
\begin{equation}
\phi_1=\phi_2=-\frac{3\pi}{4} .
\end{equation}
Moreover, with $r\rightarrow 1$ parts of eq.~(\ref{asy for p and s}) one can check that
\begin{equation}
\sqrt{\frac{2}{\pi p(r)}}\left[\cos\left(\tilde{\omega}r-\frac{3\pi}{4}\right)+i\sin
\left(\tilde{\omega}r-\frac{3\pi}{4}\right)\right]\rightarrow \text{constant}
\left(1-r\right)^{\frac{1}{2}-\frac{i\tilde{\omega}}{5}}.
\end{equation}
This indicates that the suitable solution near $r=0$ should take the form
\begin{equation}
\psi_{T,L}\left(k,r\rightarrow 0\right)=\text{constant}\left[\sqrt r J_1(\tilde{
\omega}r)+i\sqrt r Y_1(\tilde{\omega}r)\right] .
\end{equation}
Consequently, the solution to $E_{T,L}\left(k,r\to 0\right)$ is
\begin{equation}
E_{T,L}\left(k,r\to 0\right)=\text{constant}\sqrt{\frac{r}{1-r^5}}\left[\sqrt r
J_1(\tilde{\omega}r)+i\sqrt r Y_1(\tilde{\omega}r)\right] .
\end{equation}
Expanding this solution to $\mathcal {O}(r^2)$, we obtain
\begin{eqnarray}
&&\mathcal {A}_{T,L}\left(\tilde{\omega},\tilde{q}=0\right)
  =-\frac{2i}{\pi \tilde{\omega}}, \nonumber\\
&&\mathcal {B}_{T,L}\left(\tilde{\omega},\tilde{q}=0\right)
  =\frac{\left(-i+2i\gamma+\pi-2i\ln2\right)\tilde{\omega}}{2\pi} ,
\end{eqnarray}
where $\gamma$ is Euler's constant and we have omitted terms proportional to $\tilde{\omega}\ln\tilde{\omega}$ in $\mathcal{B}\left(\tilde{\omega},\tilde{q}=0\right)$ which we recognize as the contact terms. In other words, we only have interest in the leading behavior of the retarded Green's functions. Then the nonzero components of vector mesonic operator retarded Green's functions are given by
\begin{equation}
G^R_{xx}\left(\omega\right)=G^R_{yy}\left(\omega\right)
  =G^R_{zz}\left(\omega\right)=
-\frac{5\pi i+\left(5+10\ln2-10\gamma\right)}{8\pi}\mathcal {N}\omega^2 .
\end{equation}
The corresponding spectral functions for vector meson are
\begin{equation}
\mathfrak{R}_{xx}\left(\omega\right)=\mathfrak{R}_{yy}\left(\omega\right)
  =\mathfrak{R}_{zz}
\left(\omega\right)=\frac{5}{4}\mathcal {N}\omega^2 .
\end{equation}

Before concluding this section, we present related main results for the scalar mode. The appropriate solution near the boundary $r=0$ takes the form:
\begin{equation}
\varphi\left(k,r\to 0\right)=\text{constant}\sqrt{\frac{r^5}{1-r^5}}
\left[\sqrt r J_3(\tilde{\omega}r)+i\sqrt r Y_3(\tilde{\omega}r)\right].
\end{equation}
Expanding it to $\mathcal {O}(r^6)$ gives
\begin{eqnarray}
&&\mathcal {A}_\varphi\left(\tilde{\omega},\tilde{q}=0\right)=
-\frac{16i}{\pi \tilde{\omega}^3},\nonumber\\
&&\mathcal {B}_\varphi\left(\tilde{\omega},\tilde{q}=0\right)=
\frac{\left(-11i+12i\gamma+6\pi-12i\ln2\right)\tilde{\omega}^3}{288\pi},
\end{eqnarray}
where we also omit the contact terms as in the vector mode case. The result for the retarded Green's function of the scalar meson is
\begin{equation}
G^R_{\varphi\varphi}\left(\omega\right)=-\frac{6\pi i+\left(11+12\ln2-12
\gamma\right)}{48^2}\left(\frac{5}{4\pi}\right)^6\mathcal {N}^\prime\omega^6,
\end{equation}
and the spectral function for scalar meson is
\begin{equation}
\mathfrak{R}_{\varphi\varphi}\left(\omega\right)=\frac{15625}{786432\pi^5}
\mathcal{N}^\prime\omega^6 .
\end{equation}
We see that the scalar mode spectral function is proportional to $\omega^6$ while it is $\omega^2$ for vector mode. Such a scaling behavior in the high frequency limit can also be extracted from the dimensions of operators dual to $A_\mu$ and $\varphi$. This strange behavior for scalar meson signifies some non-universal characteristics of this model or this duality.

\section{Heavy quark diffusion}
\label{section6}
As the final part of this paper, we now turn to compute the heavy quark diffusion constant, which can be recognized as a comparison with light quark diffusion presented in subsection~\ref{light}. In gauge/gravity setup, heavy quark is introduced as the endpoint of an open string stretched from the boundary to the bulk in depth. Following this picture, many aspects of heavy quark moving through strongly coupled quark-gluon plasma such as jet quenching parameter \cite{Liu,Liu1,Argyres}, drag force \cite{Gubser2}, parton energy loss \cite{Herzog} have been studied (for recent reviews see \cite{Gubser1, Casalderrey-Solana}). The heavy quark diffusion constant was first computed in \cite{Casalderrey-Solana1,Gubser3,Casalderrey-Solana2} for $\mathcal{N}=4$ super-Yang-Mills plasma under the AdS/CFT technique. It has also been studied in \cite{Pang} for critical Sakai-Sugimoto model, but we will find that the behavior for heavy quark diffusion is quite different between these two models.

Before going into the detailed extraction of the heavy quark diffusion constant, we now briefly state how the diffusion of a relativistic heavy quark through the quark-gluon plasma is described by a generalized Langevin equation. We first give the physical picture from the boundary thermal field theory and then express the main formulas for heavy quark diffusion constant in terms of the quantities on the dual gravity side.

The following statements can be found in \cite{Gursoy1} and more detailed and fundamental expressions can be found in \cite{Feynman,Kleinert}. Due to the presence of the strongly-coupled quark-gluon plasma, the trajectory of a heavy quark through the plasma is believed to be described by Brownian motion. The action for an external quark moving through the plasma should take the following form:
\begin{equation}
S[X(t)]=S_0+\int d \tau X_{\mu}(\tau)\mathcal {F}^{\mu}(\tau),
\end{equation}
where $S_0$ is the free quark action, $X_{\mu}(\tau)$ is the covariant coordinate of the heavy quark and $\mathcal {F}$ represents the plasma drag force exerting on the quark. When the interesting energy scale is much smaller than the quark kinetic part, the microscopic degree of freedom of the plasma can be treated semiclassically. Then the motion of quark in the plasma is reduced to a classical generalized Langevin equation for the coordinate $X_i(t)$:
\begin{equation}
\frac{\delta S_0}{\delta X_i(t)}=\int^{+\infty}_{-\infty}d \tau \theta (\tau) C^{ij}(\tau)X_j(t-\tau)+\xi^{i}(t), \quad \langle\xi^i(t)\xi^j(t^{\prime})\rangle=A^{ij}(t-t^{\prime})\label{equation for quark}
\end{equation}
where
\begin{equation}
C^{ij}(t)=G^{ij}_{asym}(t)\equiv-i \langle[\mathcal{F}^i(t),\mathcal{F}^i(0)]\rangle,\quad
A^{ij}(t)=G^{ij}_{sym}(t)\equiv -\frac{i}{2} \langle{\{\mathcal{F}^i(t),\mathcal{F}^i(0)\}}\rangle.
\end{equation}
The first term in the first equation of eq.~(\ref{equation for quark}) is the friction part while the second one is for the random motion part.

We now introduce the correlation time $\tau_c$ and when the time scale is much larger than this correlation time, random term correlation can be approximated as follows:
\begin{equation}
A^{ij}(t-t^{\prime})\approx \kappa^{ij}\delta(t-t^{\prime}).
\end{equation}
We also define a friction function as
\begin{equation}
C^{ij}(t)\equiv \frac{d}{dt}\gamma^{ij}(t).
\end{equation}
Therefore in the large time scale (compared to the time correlation $\tau_c$) limit, eq.~(\ref{equation for quark}) reduces to the following local one
\begin{equation}
\frac{\delta S_0}{\delta X_i(t)}+\eta^{ij}\frac{d X_j(t)}{dt}=\xi^i(t),\quad \langle\xi^i(t)\xi^j(t^{\prime})\rangle=\kappa^{ij}\delta(t-t^{\prime}).
\end{equation}
In above equations, we have also defined two constant: the diffusion constant $\kappa^{ij}$ and the friction constant $\eta^{ij}$
\begin{equation}
\kappa^{ij}\equiv \lim_{\omega \to 0}G^{ij}_{sym}(\omega)=-\lim_{\omega \to 0}\frac{2T}{\omega}G^{ij}_{R}(\omega),\quad \eta ^{ij}\equiv \int^{\infty}_{0}d\tau \gamma^{ij}(\tau).
\end{equation}
In obtaining the final expression for diffusion constant, we have used the following relation for a system at equilibrium with a canonical ensemble at temperature $T$
\begin{equation}
G^{ij}_{sym}(\omega)=-\coth\frac{\omega}{2T} \text{Im}G^{ij}_{R}(\omega).
\end{equation}
And $G^{ij}_{sym}(\omega)$ is the Fourier transformation of $G^{ij}_{sym}(t)$.

Now we need to express these results in terms of variables on the dual gravity side. The only thing one needs to do is to identify the appropriate bulk field that couples to the boundary operator $\mathcal{F}$, then solve the bulk equations for this field with appropriate boundary conditions. As mentioned at the beginning of this section, a heavy quark under the viewpoint of gauge/gravity duality is described by a string with endpoint attached to the flavor brane and then extending into the bulk. If we choose the static gauge for string worldsheet (i.e., $\sigma=r$ and $\tau=t$), then the embedding coordinate is just the spacial coordinate $X^i(r,t)$. Under this trailing string picture for heavy quark, the appropriate bulk field is nothing but the quark position, i.e. the boundary value of the string embedding $X^i(r,t)$. Therefore, the correlation function of the drag force can be extracted from solving the bulk linear fluctuations around the trailing string and using the holographic method. In the rest of this section, we will carry out this calculation for the non-critical holographic QCD model.

The string worldsheet coordinate is chosen as $\tau= t$, $\sigma=r$ and we just consider a static string, i.e., $X_i$ and $X_4$ are just constants, which can be proved to be compatible with equation of motion from the Nambu-Goto action. The induced metric on the string worldsheet is
\begin{equation}
ds_{WS}^2=-\left(\frac{u_T}{R}\right)^2\frac{1}{r^2}f(r)dt^2
+\left(\frac{R}{r}\right)^2\frac{dr^2}{f(r)} .
\end{equation}
Following the procedure in \cite{Casalderrey-Solana1}, we consider the fluctuation of the static string embedding. For simplification, we just fluctuate the $X_1$ direction $X_1\rightarrow X_1+\zeta(t,r)$, and then the induced metric including fluctuation on the worldsheet is
\begin{equation}
ds^2_{WSF}=\left(\frac{u_T}{R}\right)^2\frac{1}{r^2}\left(-f(r)dt^2+
\dot{\zeta}^2 dt^2+{\zeta^\prime}^2 dr^2+2\dot{\zeta}\zeta^\prime
dt dr\right)+\left(\frac{R}{r}\right)^2\frac{dr^2}{f(r)},
\end{equation}
where the dot represents derivative with respect to $t$ and the prime means derivative with respect to $r$. The Nambu-Goto action up to quadratic order in fluctuation $\zeta$ is
\begin{equation}
S_{fl}=-\frac{u_T}{4\pi l_s^2}\int dt dr \frac{1}{r^2}\left[\frac{\dot{\zeta}^2}
{f(r)}-\frac{u_T^2}{R^4}f(r){\zeta^\prime}^2\right].
\end{equation}
In the above expression, we have omitted an infinite constant term. Now the equation of motion for fluctuation is
\begin{equation}
\frac{\ddot{\zeta}(t,r)}{r^2f(r)}-\frac{u_T^2}{R^4}\frac{\partial}{\partial r}
\left(\frac{f(r)}{r^2}\zeta^\prime(t,r)\right)=0 .
\end{equation}

As in the calculation of meson spectral functions in previous sections, we perform Fourier transformation for $\zeta(t,r)$:
\begin{equation}
\zeta\left(t,r\right)=\int\frac{d\omega}{2\pi} e^{-i\omega t}
 \zeta\left(\omega,r\right) .
\end{equation}
Under this transformation, the on-shell action and equation of motion take the forms:
\begin{eqnarray}
&&S_{\text{on-shell}}=\frac{u_T^3}{4\pi l_s^2R^4}\int\left.
 \frac{d\omega}{2\pi}\frac{f(r)}{r^2}\zeta\left(-\omega,r\right)
\partial_r\zeta\left(\omega,r\right)\right|_{r=0}^{r=1}, \\
&&\zeta^{\prime\prime}\left(\omega,r\right)+\left[\frac{f^\prime(r)}{f(r)}
 -\frac{2}{r}\right]\zeta^\prime\left(\omega,r\right)+
\frac{\tilde{\omega}^2}{f^2(r)}\zeta\left(\omega,r\right).
\label{eq for coordinate}
\end{eqnarray}
Note that the results presented here is similar to those for the vector and scalar modes, especially when we turn off the spatial momentum there. As we just want to extract out the diffusion constant following the proposition in \cite{Casalderrey-Solana1}, we here merely study the low frequency behavior of eq.~(\ref{eq for coordinate}), which is parallel to previous hydrodynamic limit study. The exponents of eq.~(\ref{eq for coordinate}) near the horizon are $\pm i\tilde{\omega}/5$ and the incoming wave boundary condition imposed at the horizon indicates that the physical solution should take the form
\begin{equation}
\zeta\left(\omega,r\right)\sim\left(1-r^5\right)^{-i\frac{\tilde{\omega}}{5}}g(r) ,
\end{equation}
here, $g(r)$ should be regular at the horizon. Then we expand $g(r)$ in order of $\tilde{\omega}$:
\begin{equation}
g(r)=g_0(r)+\tilde{\omega}g_1(r)+\ldots
\end{equation}
One can get equations obeyed by $g_0(r)$ and $g_1(r)$:
\begin{eqnarray}
&&g^{\prime\prime}_0(r)+\left[\frac{f^\prime(r)}{f(r)}-\frac{2}{r}
\right]g^\prime_0(r)=0,\\
&&g^{\prime\prime}_1(r)+\left[\frac{f^\prime(r)}{f(r)}-\frac{2}{r}\right]
g^\prime_1(r)+
\frac{2i r^4}{f(r)}g^\prime_0(r)+\frac{3ir^3}{f(r)}g_0(r)=0\label{g1} .
\end{eqnarray}
The regularity condition at the horizon $r=1$ solves $g_0(r)$ as
\begin{equation}
g_0(r)=c .
\end{equation}
Regularity together with vanishing conditions at the horizon for $g_1(r)$ also solve eq.~(\ref{g1}) as
\begin{equation}
g_1(r)=\frac{3ic}{200}\left[2\sqrt{25-10\sqrt5}\pi+5\sqrt5 \cosh^{-1}
\left(\frac{3}{2}\right)-25\ln5\right]+\frac{1}{2}icr^3+\mathcal {O}(r^4) .
\end{equation}
In the final solution for $g_1(r)$, we have expanded it to $\mathcal {O}(r^3)$ which is sufficient for calculating the diffusion constant.

The solution for $\zeta\left(\omega,r\right)$ has the required form as in eqs.~(\ref{boundary for ET}), (\ref{boundary for EL}) and (\ref{boundary for phi}):
\begin{equation}
\zeta\left(\omega,r\right)=c\left(1+\ldots\right)
   +\frac{ic\tilde{\omega}}{2}\left(r^3+\ldots\right).
\end{equation}
The retarded Green's function for the coordinate fluctuation $\zeta\left(\omega,r \right)$ is
\begin{eqnarray}
G^R(\omega)=-2\times\frac{u_T^3}{4\pi l_s^2R^4}\frac{\frac{3}{2}ic
  \tilde{\omega}}{c}
=-i\frac{12\pi R^2}{25l_s^2}T^2\omega .
\end{eqnarray}
We can first obtain the constant $\kappa$, which describes the correlation of noise in Langevin dynamics of stochastic motion\footnote{We will not go into details of this description for heavy quark moving through quark-gluon plasma, which can be found in \cite{Casalderrey-Solana1}.}, given by
\begin{eqnarray}
\kappa\equiv -\lim_{\omega \to 0}\frac{2T}{\omega} \text{Im}G^R(\omega)
=\frac{24\pi R^2}{25l_s^2}T^3 .
\end{eqnarray}
Then the heavy quark diffusion constant can be extracted from the Einstein relation for stochastic motion:
\begin{equation}
D\equiv\frac{2 T^2}{\kappa}=\frac{l_s^2}{R^2}\frac{25}{12\pi}
 \frac{1}{T} . \label{heavy quark diffusion}
\end{equation}
This result is the same as that of \cite{Casalderrey-Solana1} for $\mathcal{N}=4$ SYM plasma except for the second factor in eq.~(\ref{heavy quark diffusion}) which is model-dependent. But our result is different from that for Sakai-Sugimoto model studied in \cite{Pang} which has the scaling $\sim T^4$ in noise correlation $\kappa$. This manifests another difference between Sakai-Sugimoto model and its non-critical version.
However, we see that the light and heavy quark diffusion constants both depend on temperature as $1/T$ in this non-critical holographic QCD model, which indicates some universal features of gauge/gravity duality.

\section{Conclusion}
In this paper we investigated some properties of the chiral symmetric phase for a non-critical holographic dual model of QCD, which is described by $N_f$ pairs of D4 and $\overline{\text{D4}}$ branes in the near-horizon geometry of $N_c$ coincident D4 branes in six dimensions (we just focus on probe limit, $N_f\ll N_c$). In such a chiral symmetric phase, the $N_f$ flavor pairs extend parallel into the black hole background. Then the induced geometry on the flavor worldvolume is nearly the same as the background except one dimension less. Moreover, the embedding profile of the flavor branes is very trivial: just one pair of parallel lines in the $x_4-u$ plane. In this sense, this model is very similar to Sakai-Sugimoto model and also much simpler than the well-investigated D3/D7 system. So our studies for flavor gauge and scalar fields are similar\footnote{One main difference comes from the common factor denoted as $\mathcal {N}$ and $\mathcal {N}^\prime$ in this paper. This factor counts the degree of freedom, it is proportional to $N_c^2$ for bulk modes while $Q_cN_f$ for flavor ones.} to their bulk counterparts studied in\cite{Son,Son1,Policastro,Kovtun,Nunez,Kovtun2,Son2,Teaney,Kovtun1,Starinets,
Iqbal,Benincasa}.

We first extracted the charge susceptibility for later convenience. Fluctuations of
flavor gauge and scalar modes were elaborately analyzed, including their equations of motion, retarded Green's functions and spectral functions in hydrodynamic as well as high frequency limit. Holographic spectral function and quasi-normal mode for mesonic operator in flavored large $N_c$ gauge theory or AdS/QCD model have been studied in a huge number of papers \cite{Erdmenger1,Evans,Myers1,Mas,Dusling,Mas1,Fujita,Colangelo,Miranda,
Fujita1,Erdmenger2,Kaminski,Gulotta,Mas2,Hassanain,Hohler,Caron-Huot,Peeters,Hoyos,
Paredes,Myers,Myers2,Gursoy1}. Due to complicatedness of differential equations such as (\ref{eq:T}), (\ref{eq:L}) and (\ref{eq:scalar}), in general one cannot get analytical results and studies along this line need much numerical work. However, the physical picture underlying this study is quite general and simple. In the chiral broken phase, meson spectrum is featured with a discrete set of stable states \cite{Casero,Mazu,Mintakevich} and its spectral function is characterized by a series of $\delta$-function peaks as pointed out in \cite{Myers}. While in the chiral symmetric phase, the meson spectrum should be featured by a set of quasi-normal modes in the induced black hole metric on the flavor branes. It is regarded that this phenomenon for meson can be described by holographic meson melting \cite{Peeters,Hoyos}, but this idea is subtle \cite{Paredes}. Numerical extraction of quasi-normal modes for Sakai-Sugimoto model can be found in \cite{Evans}. We leave detailed investigations about transition from stable mesons to quasi-normal modes in this non-critical model following the approach of \cite{Myers1} for further work.

Therefore, we mainly studied two limits which allow us to use some analytical methods to get clear results. In the hydrodynamic limit, we first got analytical results for retarded Green's functions of different flavor modes. The retarded Green's functions for vector meson operators were classified according to the rotation invariance of four dimensional thermal field theory into transverse and longitudinal parts as illustrated in eq.~(\ref{retarded green function for vector}): the transverse part of vector correlator has no pole while the longitudinal one has a simple pole, which is responsible for the flavor diffusion. The scalar meson correlator has similar behavior as that of transverse part of vector meson: scaling linearly with frequency and has no pole. One apparent difference between them is their dependence on the temperature $T$: for the vector mode (both the transverse and longitudinal modes) it is linearly dependent on $T$ as shown in
eq.~(\ref{retarded green function for vector}) while for the scalar meson $T^6$-dependent as shown in eq.~(\ref{retarded green function for scalar}). These characteristics are consistent with previous studies both for bulk fields and flavor fields. Spectral function is then just the byproduct of retarded Green's function. Although these mesonic correlation functions and spectral functions have the same structure as those of other holographic models, either from bottom-up approach or top-down intersecting D-brane systems, but one main difference between them is that their scaling dependence on the temperature and frequency, which should be recognized as model-dependent.

We then extracted the light quark diffusion constant from the simple pole of longitudinal part of vector mode correlator. This lowest quasi-normal frequency stands for the dispersion relation $\omega=-iDq^2$ where $D$ is just the diffusion constant. We also used another two different means to compute this constant and performed the cross check. One of the methods is using Green-Kubo formula and the other one is using transport coefficient formula derived in \cite{Kovtun,Starinets,Iqbal} under black hole membrane paradigm. The same results by using different methods indicate some universal features of gauge/gravity duality. Another limit case which allows for analytical investigation is the high frequency or zero temperature limit. We first inverted these second order differential equations to standard Schr$\ddot{\text{o}}$dinger types and then used WKB approximation to solve them. We found that the spectral functions are not dependent on temperature which is just the trivial result of zero temperature limit and their dependence on frequency can be simply extracted from dimensional analysis for corresponding dual operators as pointed out in \cite{Myers}.

In the final part of this paper we focused on the heavy quark diffusion constant in this model.  This constant plays an important role in heavy ion physics that can be studied at RHIC and LHC experiments. We followed the prescription in \cite{Casalderrey-Solana1} to concisely calculate this constant and found similar behavior with previous studies \cite{Casalderrey-Solana1,Gubser3,Casalderrey-Solana2} for $\mathcal{N}=4$ super-Yang-Mills plasma but different from \cite{Pang} for Sakai-Sugimoto model. This result may signal one main difference between Sakai-Sugimoto model and its non-critical version.

There are some other interesting problems which are not covered in this work but worth of investigation in future work. The first one is about the relationship between stable meson state in chiral broken phase and its quasi-normal mode counterpart in chiral symmetric phase. In this kind of model, the flavor branes and color branes are transversely intersecting so that the bare quark mass is exactly zero. So, introducing
quark mass and probing its effect on spectral function or meson melting in the process of chiral phase transition is interesting. When turning on background for U(1) gauge field to model external electromagnetic fields and chemical potential, the calculations for correlation functions are more difficult due to coupling of different flavor modes. Studies along this line can be found in \cite{Erdmenger1,Mas,Mas1,Caron-Huot,Erdmenger2,
Kaminski}. Generalizing these studies to other holographic systems also seems interesting. We have focused on the probe limit in this paper, but the supergravity background in six dimension taking into the flavor backreaction has also been constructed in \cite{Kuperstein}, so studying meson spectral functions beyond probe limit in this non-critical holographic QCD model also deserves explorations.

\acknowledgments{YYB thanks Zhen-Hua Zhang for useful discussions. JMY thanks JSPS for the invitation fellowship (S-11028) and the particle physics group of Tohoku University for their hospitality. This work was supported in part by NSFC (Nos.~10821504,10725526).}

\end{document}